\documentclass[aps,prl,twocolumn,preprintnumbers,amsmath,amssymb,superscriptaddress]{revtex4-2}
\usepackage[utf8]{inputenc}
\usepackage[english]{babel}
\usepackage[T1]{fontenc}
\usepackage{amsmath}
\usepackage{hyperref}
\usepackage{amsmath}
\usepackage{amssymb}
\usepackage{physics}
\usepackage{graphicx}
\usepackage{booktabs}

\usepackage{booktabs,multirow}

\hypersetup{
    colorlinks=true,
    linkcolor=blue,
    filecolor=blue,      
    urlcolor=blue,
    citecolor=blue,
}

\begin{document}
\title{Effective temperature in approximate quantum many-body states}

\author{Yu-Qin Chen}
\email{yqchen@gscaep.ac.cn}
\affiliation{Graduate School of China Academy of Engineering Physics, Beijing 100193, China}

\author{Shi-Xin Zhang}
\email{shixinzhang@iphy.ac.cn}
\affiliation{Institute of Physics, Chinese Academy of Sciences, Beijing 100190, China}

\begin{abstract}

In the pursuit of numerically identifying the ground state of quantum many-body systems, approximate quantum wavefunction ansatzes are commonly employed. This study focuses on the spectral decomposition of these approximate quantum many-body states into exact eigenstates of the target Hamiltonian.  The energy spectral decomposition could reflect the intricate physics at the interplay between quantum systems and numerical algorithms. Here we examine various parameterized wavefunction ansatzes constructed from neural networks, tensor networks, and quantum circuits, employing differentiable programming to numerically approximate ground states and imaginary-time evolved states. Our findings reveal a consistent exponential decay pattern in the spectral contributions of approximate quantum states across different ansatzes, optimization objectives, and quantum systems, characterized by small decay rates denoted as inverse effective temperatures. The effective temperature is related to ansatz expressiveness and accuracy and shows phase transition behaviors in learning imaginary-time evolved states. The universal picture and unique features suggest the significance and potential of the effective temperature in characterizing approximate quantum states.
\end{abstract}

\maketitle
\noindent{\large{\textbf{Introduction}}} 

\noindent Understanding the ground state properties of quantum many-body systems is a central challenge in modern physics, with broad implications ranging from fundamental principles of quantum complexity theories to the design of novel materials and quantum technologies \cite{Kitaev2002a_z,  Kempe2006_z,  Zheng2017, Lee2023}. The exponentially large Hilbert space with the system size often precludes analytical solutions and exact numerical methods, necessitating the development of approximate numerical methods. Among these, the use of approximate quantum wavefunction ansatzes has proven to be powerful for its good trade-off between expressivity and complexity.

In this work, we investigate a profound and previously unexplored aspect of approximate quantum states: the effective temperature that characterizes their spectral properties. The concept of temperature arises from statistical mechanics \cite{Greiner1995}. Here, we apply this concept to quantum many-body pure states, specifically to the approximate states obtained through optimizing different ansatzes. We investigate the effective temperature and its dynamics during variational training, using both energy expectation and fidelity as objectives.

Our work reveals a surprising universality in the effective temperatures of approximate quantum ground states, irrespective of the ansatz structure, objective function, training wellness, or underlying physical system. By decomposing the approximate states into the exact eigenstates of the system Hamiltonian, we observe a spectrum extending to the high-energy end with exponential decay of small decay factors, interpreted as inverse effective temperatures. 
The effective temperature also shows an intriguing two-stage behavior when approximating pure states of finite ``temperature".
The universal pattern could play an important role in understanding complex quantum systems or designing new classical and quantum algorithms. The effective temperature offers a new lens to assess the accuracy and reliability of different variational ansatzes and algorithms, potentially guiding the design of more efficient methods.

\begin{figure*}[t]\centering
	\includegraphics[width=0.93\textwidth]{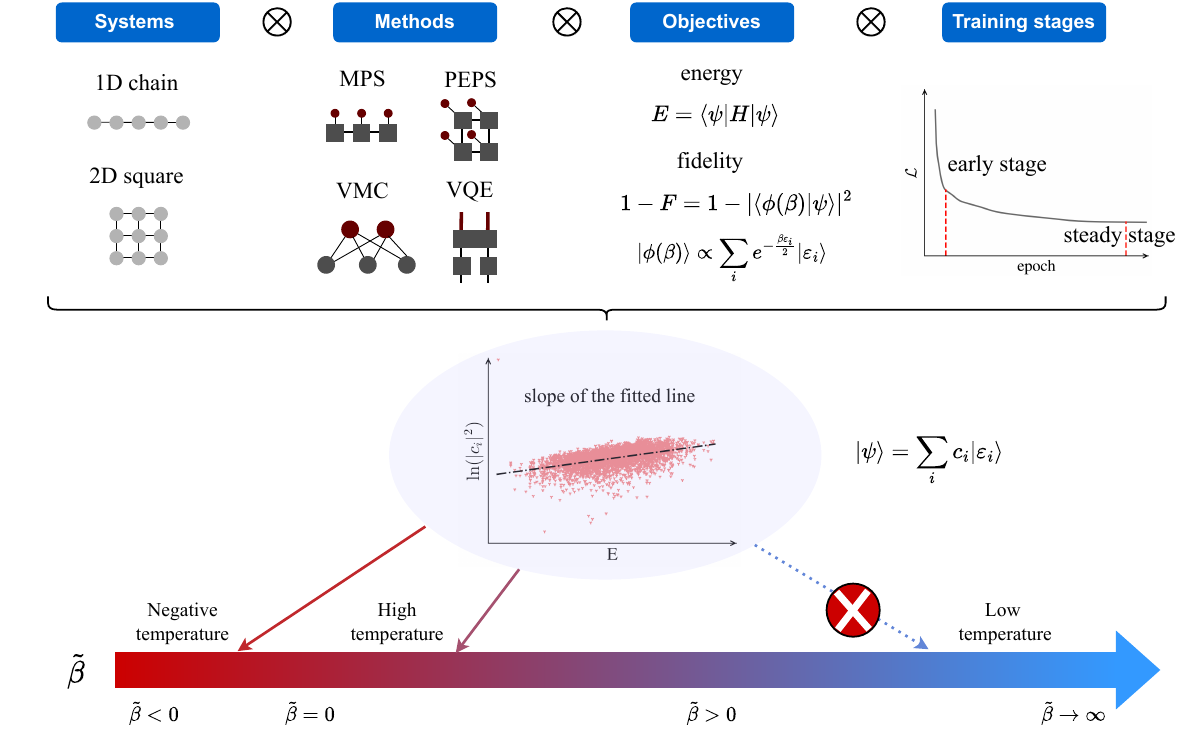}
	\caption{\textbf{Sketch of the effective temperature as a metric characterizing different systems and numerical methods.} The effective temperature ($1/\tilde{\beta}$) can be extracted from the estimated slope of the approximate quantum state spectral decomposition. The metric varies with different systems, methods, objectives, and training wellness. However, universal features remain the same and are valuable in diagnosing the capacity of different numerical methods. In general, $\tilde{\beta}$ is small and can be even negative, indicating a nearly flat excited-states spectrum of approximate ground states.  }
\label{fig:workflow}
\end{figure*}

~\\

\noindent{\large{\textbf{Setups}}}

\noindent A wavefunction ansatz comprises a family of parameterized quantum states as $\vert \psi(\theta)\rangle$ with tunable parameters $\theta$. Such quantum states usually admit a compact and scalable representation on classical computers or quantum computers to circumvent the challenge of the exponentially large Hilbert space for generic quantum many-body systems. The approximate state within the manifold of a given ansatz can be obtained through variational optimization \cite{Wu2024} $\vert \psi\rangle = \text{min}_\theta \mathcal{L}(\vert \psi(\theta)\rangle)$, where $\mathcal{L}$ is the objective function built on top of the wavefunction. Specifically, 
\begin{equation}
{\mathcal{L}_E = \langle \psi(\theta)\vert H\vert \psi(\theta)\rangle}
\end{equation}
is often utilized to obtain the ground state of the system Hamiltonian $H$. The Hamiltonian has an exact energy spectral decomposition as $H = \sum_{i=0} \varepsilon_i \vert \varepsilon_i\rangle \langle \varepsilon_i\vert $ with $\vert \varepsilon_0\rangle$ being the exact ground state. Additionally, we also frequently utilize the fidelity objective
\begin{equation}
\mathcal{L}_F = 1-\vert\langle \varepsilon_0\vert \psi(\theta)\rangle \vert^2 
\end{equation}
for numerically approaching ground states. Throughout this work, we perform optimization via gradient descent, namely, the variational parameters are updated in each step according to $\Delta \theta \propto \frac{\partial \mathcal{L}}{\partial \theta}$, where the gradients can be efficiently evaluated using automatic differentiation \cite{Rumelhart1986a, Bartholomew-Biggs2000a, Liao2019, Zhang2019b}.

For an approximate state $\vert \psi\rangle$, we decompose it on the basis of the exact eigenstates $\vert \varepsilon_i\rangle$ of Hamiltonian $H$:
\begin{equation}
    \vert \psi\rangle = \sum_i c_i \vert \varepsilon_i\rangle.
\end{equation}
The spectral decomposition coefficients $\vert c_i\vert^2$ are the central focus of our study. We find that the pairs of $(\varepsilon_i, \vert c_i\vert^2)$ roughly follow an exponential decay for the approximate ground state, analogous with Boltzmann distribution,  
\begin{equation}
\vert c_i\vert^2\propto e^{-\tilde{\beta} \varepsilon_i},
\end{equation}
where we denote the decay factor $\tilde{\beta}$ as the inverse effective temperature for the pure state $\vert \psi\rangle$ under investigation.

To gain deeper insights into the spectrum pattern of approximate quantum states, we evaluate additional target states beyond the ground state $\vert \varepsilon_0\rangle$. This set of states is directly parameterized with an inverse effective temperature $\beta$ as
\begin{equation}
    \vert \phi(\beta)\rangle =\frac{1}{Z(\beta)}\sum_i e^{-\frac{\beta \varepsilon_i}{2}}\vert \varepsilon_i\rangle,
\end{equation}
where $Z(\beta)=\sqrt{\sum_i e^{-\beta \varepsilon_i}}$ is the normalization factor. These states can be understood as imaginary-time evolved states with time $\beta$ from the initial state $\vert \phi(0)\rangle=\frac{1}{Z(0)}\sum_i \vert \varepsilon_i\rangle$. Therefore, we call the target states imaginary-time evolved states (ITES). Exact ITES by default admits an exponential decay for spectral decomposition with inverse temperature $\beta$. We investigate how spectral properties change for approximate ITES $\vert \psi\rangle$ which are obtained by optimizing the fidelity objective $\mathcal{L}_{F}(\beta)=1-\vert\langle \phi(\beta)\vert \psi(\theta)\rangle \vert^2 $. This objective reduces to the ground state target in the $\beta\rightarrow\infty$ limit. We find that the approximate ITES spectrum shows an evident two-stage behavior for all ansatzes: $\tilde{\beta}$ from approximate ITES successfully matches $\beta$ for small $\beta$ (high-temperature regime) and shows deviation $\tilde{\beta}<\beta$ for large $\beta$ (low-temperature regime). More importantly, the transition point $\beta^*$ characterizes the lowest effective temperature a given ansatz can reach and implies the intrinsic power of the corresponding ansatz. Notably, the conclusion remains qualitatively the same when the coefficient for each eigenstate in ITES acquires an extra random phase for generality, i.e. $\vert \phi(0)\rangle = \frac{1}{Z(0)}\sum_i e^{i\omega_i}\vert \varepsilon_i\rangle$.

In this work, we employ various quantum wavefunction ansatzes to demonstrate the universality of our findings, including (1) tensor-network based ansatz including matrix product states (MPS) \cite{White1992, Schollwock2011,Stoudenmire2012} and projected entangled-pair states (PEPS) \cite{Verstraete2004, Verstraete2004a}, (2) neural quantum states (NQS) built on top of neural networks \cite{Carleo2017, Deng2017d}, (3) output states from variational quantum circuits \cite{Cerezo2021, Bharti2021} such as variational quantum eigensolvers (VQE), \cite{Peruzzo2014, Tilly2022} and (4) vector state ansatz (VEC) where each wavefunction component is directly modeled as a trainable parameter. We also comment on the relevance of our results to quantum approximate optimization algorithms (QAOA) \cite{Farhi2014}. We apply these ansatzes to different system Hamiltonians on different lattice geometries and optimize both objectives $\mathcal{L}_E$ and $\mathcal{L}_F$. The main findings of this work are summarized in Fig. \ref{fig:workflow}.

\begin{figure}[t]\centering
	\includegraphics[width=0.49\textwidth]{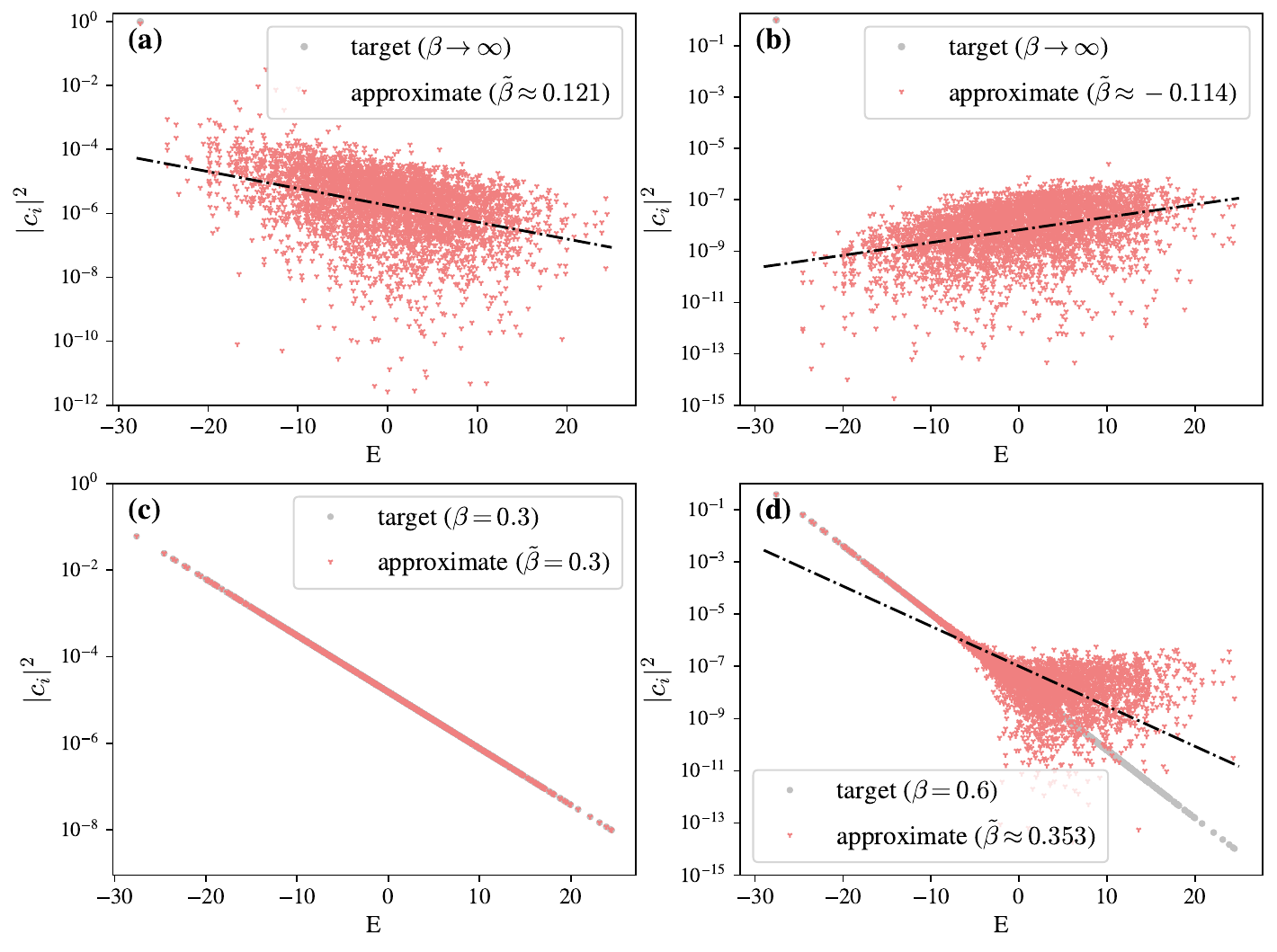}
	\caption{\textbf{Spectral decomposition of approximate states.} The 2D XXZ model on $4\times 3$ square lattice with $J_x=J_y=1, J_z=0.8$ and MPS ansatz with bond dimension $\chi=32$ are employed. The optimization objective is the infidelity between approximate states and the target states. The target states are chosen as the ground states for (a)(b) and imaginary-time evolved states $\vert \phi(\beta)\rangle$ with $\beta=0.3$ (c) and $\beta=0.6$ (d). Overlaps with different energy eigenstates are depicted as scatter plots in red and gray for approximate and target states, respectively. The optimization is at the early stage for (a) and converged for (b)(c)(d) at the steady stage toward the target state. The logarithmic overlaps with excited eigenstates show a linear pattern during training with varying fitted slope $\tilde{\beta}$, aka. inverse effective temperature. The fitted line is shown in dashdot.  When the target state is ITES, there is a phase transition in the spectrum patterns of the approximate state. For small $\beta$ in (c), the optimized approximate state exhibits a near-perfect correspondence with the target state overlap, with a fitted slope $\tilde{\beta}$ matching $\beta = 0.3$. Conversely, for large $\beta$ in (d), the overlaps from approximate states exhibit distinct behaviors -- an exponential decay in the lower energy regime and a plateau in the higher energy regime. This behavior leads to a poor linear fit, characterized by a deviating slope $\tilde{\beta}<\beta$. }
\label{fig:mps32_overlap}
\end{figure}

~\\

\noindent{\large{\textbf{Results for approximate ground states}}}

\noindent Since producing the exact spectrum requires a full diagonalization of the Hamiltonian, our numerical study is limited to small systems up to size $L=16$. We use {\sf TensorCircuit-NG}~\cite{*[{ }] [{. https://github.com/tensorcircuit/tensorcircuit-ng.}] Zhang2022} for the numerical simulation of variational training and {\sf QuSpin}~\cite{10.21468/SciPostPhys.2.1.003} for the exact diagonalization of the Hamiltonian within full Hilbert space or specific charge sectors.

We focus on the two-dimensional XXZ model on a square lattice with spin-$1/2$ degrees of freedom:
\begin{equation}
H = \sum_{\langle ij\rangle} J_x X_iX_j + J_y Y_iY_j + J_z Z_iZ_j,
\end{equation}
where $X(Y, Z)_i$ are Pauli matrices on lattice site $i$ and $\langle ij\rangle $ denotes the nearest neighbor sites. The phase diagram of this model, as given in Ref. \cite{Yunoki2002}, features $J_{z}^c=1$ that separates the spin-flipping phase and the antiferromagnetic phase when $J_x=J_y=1$. We impose periodic boundary conditions throughout this work. For $4\times 4$ lattice, we focus on the half-filling charge sector for spectral decomposition while the variational wavefunction ansatzes are still expressed and optimized in the full Hilbert space for simplicity. For  $4\times 3$ lattice, we utilize the full spectral decomposition for most of the ansatzes while restricting to the half-filling sector for VQE states (see SM for details).

\begin{figure}[t]\centering
	\includegraphics[width=0.48\textwidth]{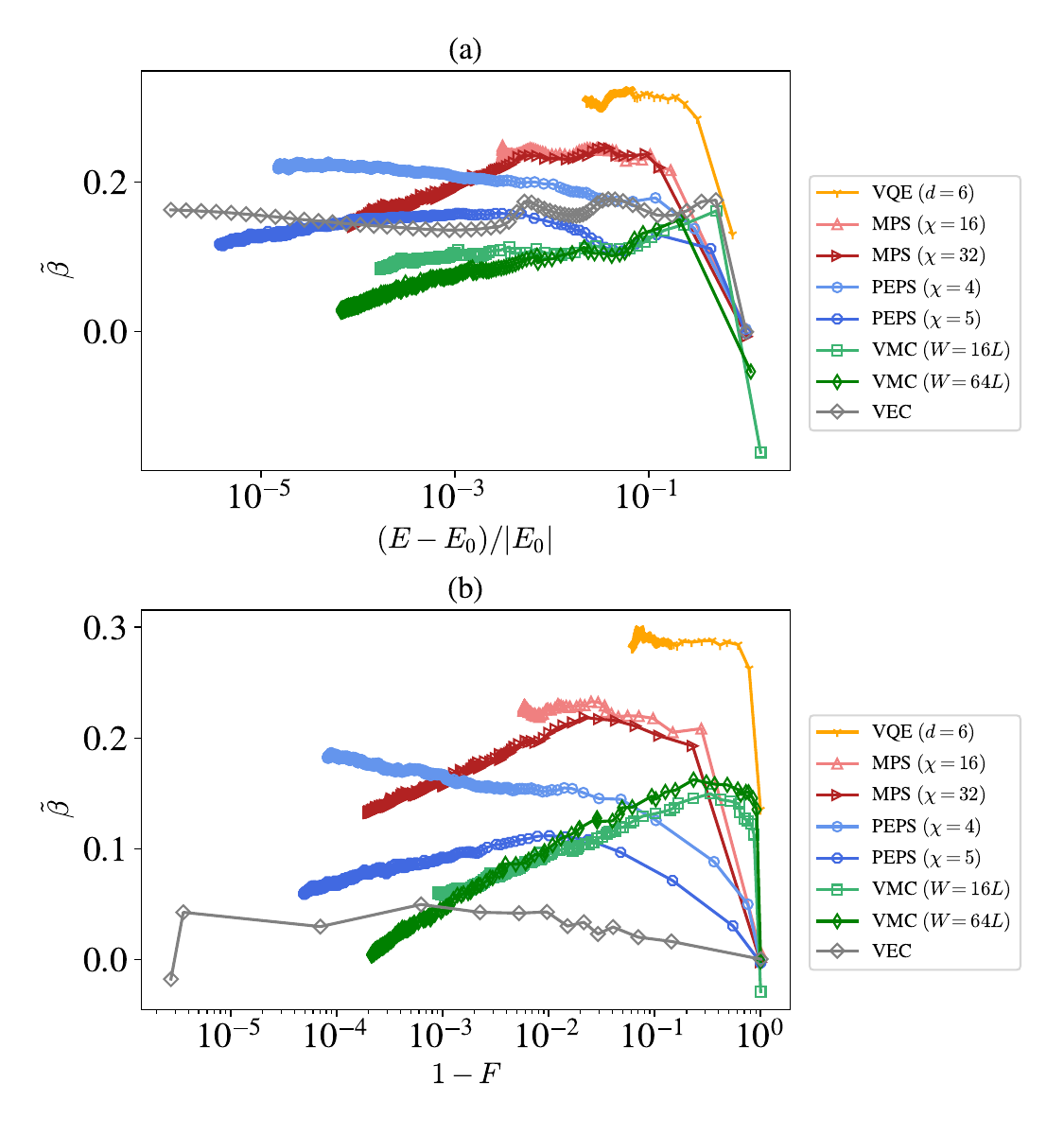}
	\caption{\textbf{Effective temperature during variational training targeting the ground states.} The objectives are energy $\mathcal{L}_E$ (a) and fidelity $\mathcal{L}_F$ (b), respectively. The results are obtained from the 2D XXZ model on $4 \times 4 $ square lattice with $J_x=J_y=1$, $J_z=0.8$. During training, points move to the left toward higher accuracy and $\tilde{\beta}$ first increases and then decreases in general.}
\label{fig:beta_training_44_110.8}
\end{figure}

Using different ansatzes and variational optimization against $\mathcal{L}_E$ or $\mathcal{L}_F$, we obtain different approximate ground states. The typical spectral decomposition patterns $(\varepsilon_i, |c_i|^2 )$ of these states are shown in Fig. \ref{fig:mps32_overlap} (a)(b). (See SM for more spectral decomposition results from other variational ansatzes.) We find a robust exponential relation in the spectral decomposition across different ansatzes and training stages. The exponential decay factor $\tilde{\beta}$ extracted from the fitting plays the role of inverse effective temperature. The training dynamics of $\tilde{\beta}$ for different ansatzes and objectives is shown in Fig. \ref{fig:beta_training_44_110.8}. We find that the effective temperature $1/\tilde{\beta}$ is high during training and this is particularly true for optimized approximate ground states where $\tilde{\beta}<0.3$ in most cases. Such a small $\tilde{\beta}$ contrasts sharply with imaginary-time evolution based approaches where a large $\tilde{\beta}$ is expected.  During variational training, a general trend emerges where $\tilde{\beta}$ first increases and then decreases, implying an upper bound for the possible $\tilde{\beta}$ of a given ansatz. Besides,  training dynamics of $\tilde{\beta}$ show similar patterns for both objectives within the same ansatz while  $\tilde{\beta}$ for optimizing energy objectives $\mathcal{L}_E$ is typically higher due to the unequal penalty on different excited states. Conversely, fidelity objective equally penalizes all excited states and the finite $\tilde{\beta}$ in this case is attributed to the physical prior encoded in the ansatz structures. This understanding is validated by the near-zero $\tilde{\beta}$ for VEC ansatz, which contains the least physical prior in the structure. Furthermore, the $\tilde{\beta}$ trends are more closely aligned within the same family of ansatzes, reflecting its potential power to diagnose intrinsic properties in these ansatzes.
The universality and validness of the results are also confirmed with other system sizes, models, and lattice geometries (see SM for details). 

~\\

\noindent{\large{\textbf{Results for approximate imaginary-time evolved states}}}

\noindent To better understand the spectrum pattern and the underlying mechanism of effective temperature in approximate ground states, we evaluate approximate ITES whose targets have strictly exponential decay spectrum patterns. For different target $\beta$s, we identify a two-stage behavior with a putative phase transition or crossover in between. For the high-temperature regime of small $\beta$, the approximate ITES conforms to the strict exponential decay spectrum pattern with $\tilde{\beta}=\beta$, as shown in Fig. \ref{fig:mps32_overlap} (c). On the contrary, for the low-temperature regime of large $\beta$, the spectral decomposition follows the exponential decay only for the low-energy part, while a nearly flat spectrum emerges from high-energy contributions.  If we attempt to extract a single exponential decay factor $\tilde{\beta}$ in this case, its value is significantly smaller than the exact $\beta$, as shown in Fig. \ref{fig:mps32_overlap} (d). It is worth noting that the nearly flat spectrum deviation is not due to numerical machine precision, as points $|c_i|^2$ of the same magnitude conform to the exponential decay for small $\beta$ but fail to do so for large $\beta$  (see SM for details). 

\begin{figure}[t]\centering
	\includegraphics[width=0.49\textwidth]{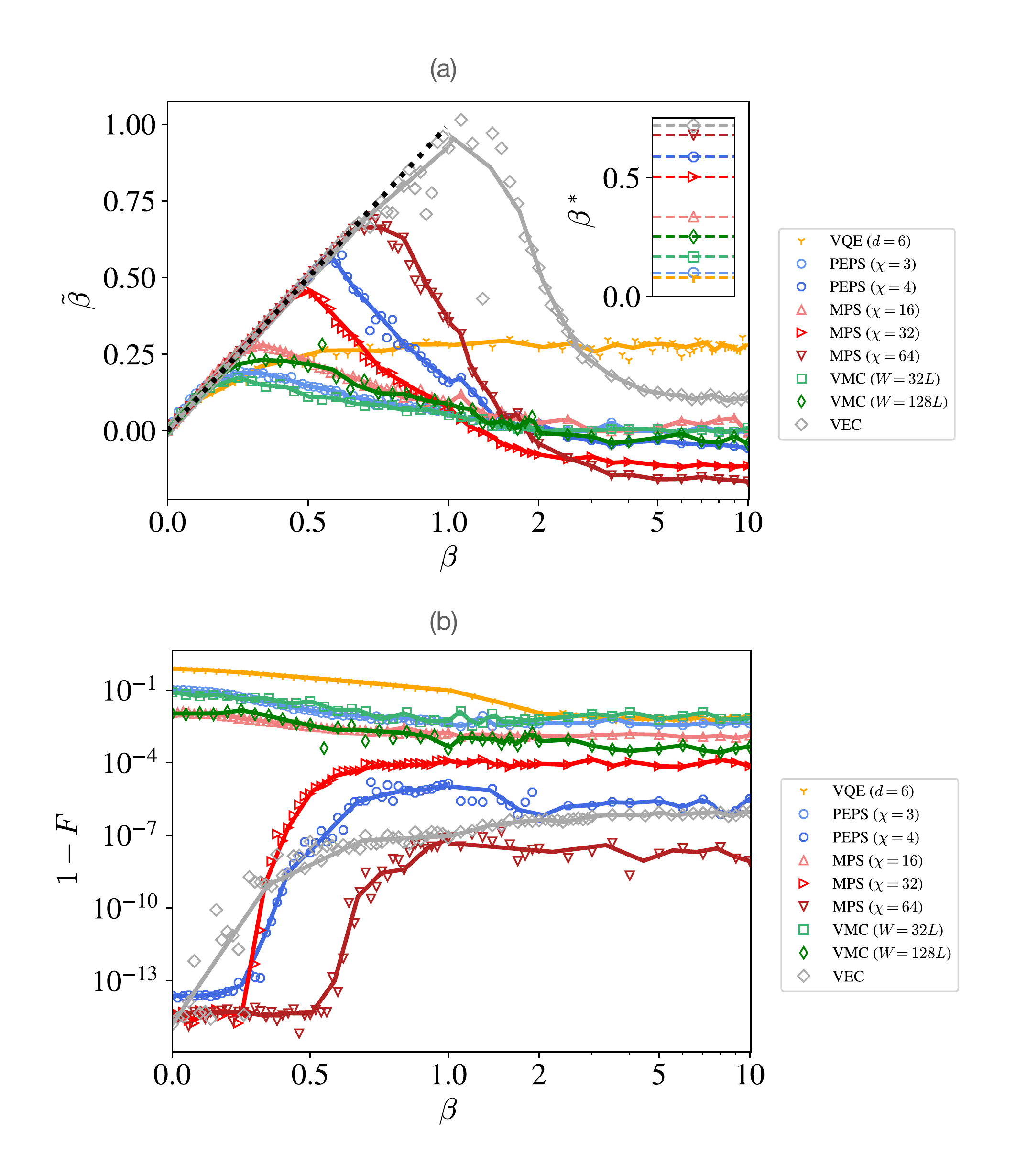}
	\caption{\textbf{Effective temperature and fidelity of approximate ITES with different $\beta$s}. The results are obtained from the 2D XXZ model on $4 \times 3 $ square lattice with $J_x=J_y=1$, $J_z=0.8$. (a) The inverse effective temperature $\tilde{\beta}$ shows a two-stage behavior with varying $\beta$. $\beta^*$ (insets) marks the deviation of $\tilde{\beta}$ from target $\beta$, separating the two stages. (b) The fidelity of approximate ITES with different $\beta$s. For high-accuracy methods, the fidelity decreases with increasing $\beta$ due to the increased optimization hardness. For low-accuracy methods, the fidelity increases with increasing $\beta$ as the target states are less entangled and are better suited for the limited ansatz. }
\label{fig:beta_beta}
\end{figure}

The results of $\tilde{\beta}$ of approximate ITES for different target $\beta$s are shown in Fig. \ref{fig:beta_beta} (a), where the two-stage behavior is evident. For each ansatz, there is a transition point $\beta^*$ when $\tilde{\beta}$ begins to deviate from $\beta$ and the associated spectrum pattern moves from Fig. \ref{fig:mps32_overlap} (c) to Fig. \ref{fig:mps32_overlap} (d). $\beta^*$ is a figure of merit as it marks the critical temperature that the ansatz can accurately represent and sets an upper bound on $\tilde{\beta}$ that an ansatz can reach. More importantly, $\beta^*$ is also correlated with the expressive capacity of the ansatz, as a higher $\beta^*$ generally implies a better fidelity. In the low temperature limit $\beta\rightarrow \infty$, the results for approximate ITES are reduced to the results for approximate ground states by optimizing $\mathcal{L}_f$.

We further remark on the fidelity results shown in Fig. \ref{fig:beta_beta} (b) with varying methods and $\beta$s. For low-accuracy ansatzes, the fidelity gets improved with increasing $\beta$ as $\vert \phi(\beta)\rangle$ of larger $\beta$ is closer to the ground state and has smaller entanglement. Such less entangled states can be better suited in ansatzes of limited expressiveness. Conversely, for high-accuracy ansatzes such as tensor network ansatz with large bond dimensions or VEC, the expressive capacity is sufficient for targeting any ITES, and the bottleneck is the optimization difficulty. Since the number of optimization steps is fixed in our simulation, a slower optimization speed results in a worse final fidelity with increasing $\beta$. In other words, when the target is closer to the ground state manifold, the optimization landscape is harder to navigate (see quantitative analysis for the expressiveness and optimization hardness in the SM).

~\\

\noindent{\large{\textbf{Discussions and Conclusion}}} 

\noindent Some hints of the general picture presented in this work are previously reported in QAOA \cite{Diez-Valle2023a, Lotshaw2023} and MPS cases \cite{Silvester2024}, where the optimization of $\mathcal{L}_E$ is considered. In the QAOA case, the output states from the QAOA circuits were found to approximate Gibbs distribution with $\tilde{\beta}\approx 0.2\sim 0.3$ \cite{Diez-Valle2023a}, consistent with the findings in this study, which can be regarded as a special limit of the VQE ansatz. In the MPS case, a nearly flat spectrum is identified \cite{Silvester2024}. Given that the density of states was effectively considered with Gaussian broadening in their work, the spectrum appeared flatter as the exponential decay of small positive $\tilde{\beta}$ was compensated with the high density of states in the middle-energy regime.

The effective temperature has profound implications for understanding numerical methods and many-body systems. One practical implication of such a high-temperature tail in the spectrum is mentioned in Ref. \cite{Silvester2024}, where sampling based energy variance estimation becomes more fragile than expected. The variance $H^2$ is just an example for operator $f(H)$, with expectation $\langle \psi\vert f(H)\vert \psi\rangle=\sum_i |c_i|^2 f(\varepsilon_i)$. The operator expectation is highly sensitive to small high-energy components when $f(\varepsilon)$ is significantly larger for high-energy inputs. The effective temperature captures finer details than fidelity as demonstrated in Fig. \ref{fig:beta_beta}: while the fidelity remains the same for larger $\beta$, the effective temperature continues to vary. This quantity also has strong relevance to the expressive capacity of ansatzes as $\beta^*$ values differ for different ansatzes, which further confirms that the two-stage behavior in approximating ITES is not a numerical precision issue. Consequently, the metric and the methodology explored in this work serve as a guiding principle to help design better approximate quantum ansatzes and variational algorithms.

The effective temperature metric opens up several intriguing future research directions. We can quantitatively investigate the scaling of effective temperature $\tilde{\beta}$ and its critical value $\tilde{\beta}^*$ with respect to varying system sizes, Hamiltonian parameters, and ansatz structure parameters, to gain deeper insights into the physics of the system and the intrinsic patterns in numerical methods. It is also interesting to explore whether some approximation methods can be developed to estimate the effective temperature beyond exact diagonalization sizes. The ITES proposed in this work are also of independent academic interest, as they provide a platform to explore the trade-off between expressiveness and optimization hardness, as well as host a putative phase transition at $\beta^*$, the nature and universal behavior of which merit future exploration. It is also crucial to validate the general picture of spectrum patterns on more ansatzes including mean-field ansatz, physics-inspired ansatz \cite{Yokoyama1987} and advanced hybrid variational ansatzes \cite{Zhang2021b, Yuan2021}, and more models including integrable systems, many-body localized systems and fermionic systems.

In this work, we have identified a universal pattern in the spectrum of approximate ground states. We define the effective temperature $\tilde{\beta}$ to characterize the exponential decay in the spectrum and investigate the behavior of $\tilde{\beta}$ with different systems, ansatzes, objectives, and training steps. To better understand the underlying mechanism for this metric, we further propose ITES targets and identify the two-stage behaviors in approximating ITES with the figure of merit critical inverse effective temperature $\beta^*$. We believe that the universal picture presented here provides a fresh and powerful perspective on benchmarking numerical algorithms and understanding quantum many-body systems.

~\\

\noindent{\large{\textbf{Methods}}}

\noindent We describe the details of different quantum many-body ansatzes employed in this work. Additional hyperparameters in terms of optimization can be found in Supplemental Materials.

\begin{figure}[t]\centering
	\includegraphics[width=0.37\textwidth]{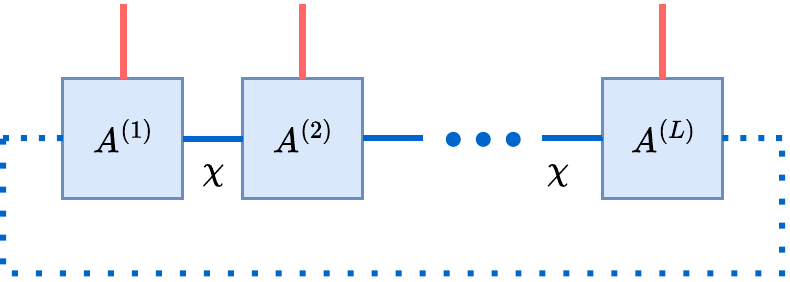}
	\caption{\textbf{The tensor network graphic representation for the periodic MPS ansatz used in this work}. Each blue square corresponds to a tensor of shape $(\chi, \chi, 2)$. The red legs correspond to physical indices of dimension $2$ and all blue legs are virtual bonds of dimension $\chi$. The dashed blue line is for the bond connecting the tensor on the last site and the first site for periodic boundary conditions.}
\label{fig:mps_ansatz}
\end{figure}

{\bf Matrix product states.} We use MPS with periodic boundary conditions. The amplitude for the ansatz is given as
\begin{equation}
    \langle i_1i_2\cdots i_L \vert \psi\rangle = \mathrm{Tr}(\prod_{j=1}^L A_{i_j}^{(j)}),\label{eq:mps_ansatz}
\end{equation}
where $i_j=0$ or $1$ represents the computational basis and $A^{(j)}$ are $L$ different tensors with shape $(\chi, \chi, 2)$. $\chi$ is the bond dimension controlling the expressiveness of the MPS ansatz. For $\chi = 2^{L/2}$, the ansatz can be an exact representation for any wavefunction in principle even with open boundary conditions. The tensor network graphic representation for the ansatz in Eq. \eqref{eq:mps_ansatz} is shown in Fig. \ref{fig:mps_ansatz}. Note that we don't explicitly impose the normalization conditions by canonicalization but leave the MPS ansatz unnormalized. The Hamiltonian expectation is evaluated as 
\begin{equation}
  E=\frac{\langle \psi\vert H\vert\psi\rangle}{\langle \psi\vert\psi\rangle}, \label{eq:mps_energy}
\end{equation}
where normalization is explicitly accounted for. For 2D systems, we use the natural order from site $1$ to $L$ to encode the system sites into 1D MPS. Since we need to consider the full spectrum properties in the calculation, the system size accessible is limited to $L=16$. Therefore, we don't rely on the density matrix renormalization group optimization algorithm. Instead, we directly obtain the full wavefunction by contracting the MPS according to Eq. \eqref{eq:mps_ansatz} and directly evaluate the Hamiltonian expectation by sparse matrix-vector multiplication where the objective can then be differentiated and variational optimized via gradient descent. For variational optimization, each element in $A^{(j)}$ is randomly initialized according to the Gaussian distribution of mean $0$ and standard deviation $0.1$. The total number of variational parameters for MPS ansatz is $2L\chi^2$.

\begin{figure}[t]\centering
	\includegraphics[width=0.43\textwidth]{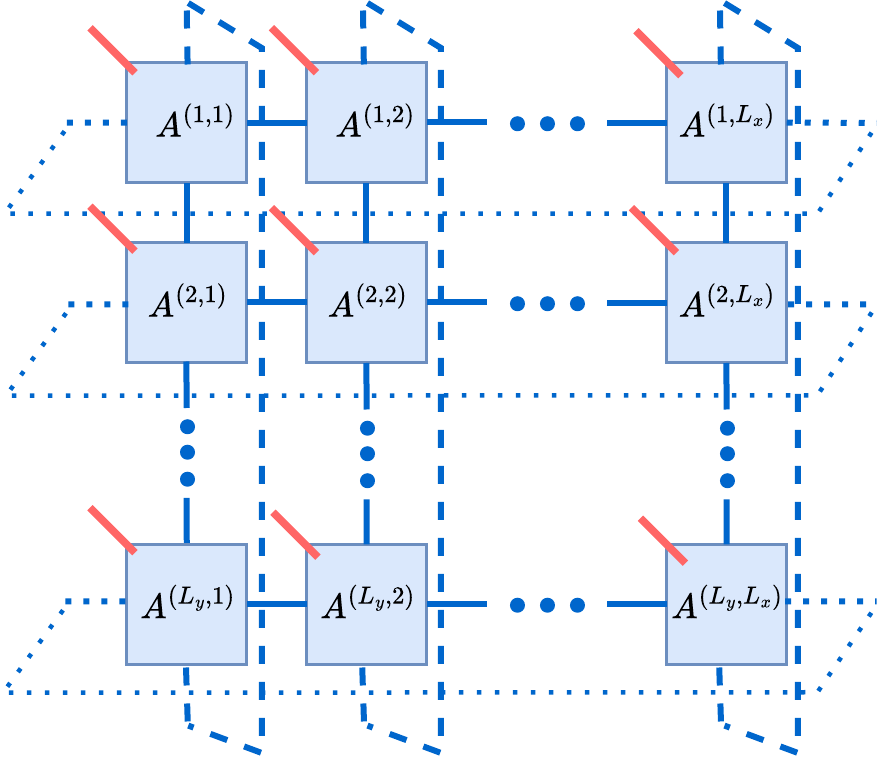}
	\caption{\textbf{The tensor network graphic representation for the two dimensional PEPS ansatz with periodic boundary conditions used in this work.} Each blue square correspond to a tensor of shape $(\chi, \chi, \chi, \chi, 2)$. The red legs correspond to physical indices of dimension $2$ on each site and all blue legs are virtual bonds of dimension $\chi$. Dashed blue lines are for the periodic boundary connections on both x and y dimensions.}
\label{fig:peps_ansatz}
\end{figure}

{\bf Projected entangled-pair states.} The tensor network graphic representation of the PEPS ansatz with periodic boundary conditions is shown in Fig. \ref{fig:peps_ansatz}.
Similarly, the ansatz is unnormalized by default. During optimization, $L=L_x\times L_y$ tensors with shape $(\chi, \chi, \chi,\chi, 2)$ are randomly initialized with each element following Gaussian distribution of mean $0$ and standard deviation $0.1$. Henceforth, the total number of variational parameters for PEPS is $2L\chi^4$. In the numerical calculation, we first contract the entire PEPS tensor network with physical legs open to directly obtain the full unnormalized wavefunction since the system size is small. We can then compute fidelity or Hamiltonian expectation directly and integrate the whole computation graph with automatic differentiation and variational optimization.

\begin{figure}[t]\centering
	\includegraphics[width=0.37\textwidth]{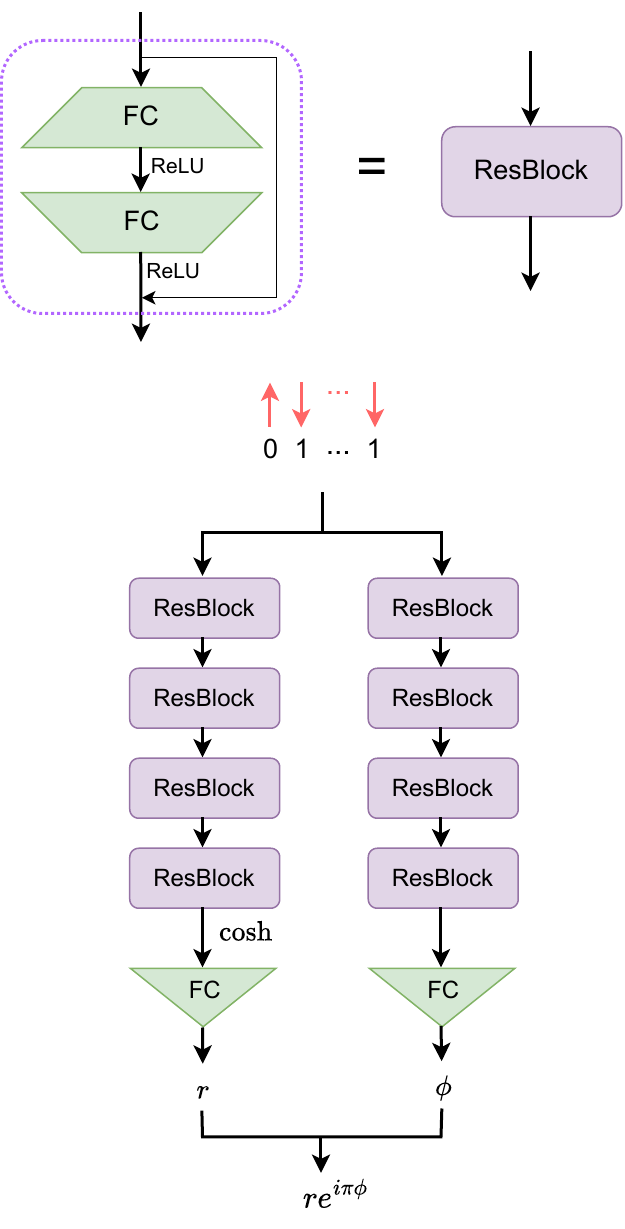}
	\caption{\textbf{Neural network architecture for the NQS used in this work.} The wavefunction amplitude and phase are evaluated separately by going through a stack of residue blocks (ResBlock). In each ResBlock, the input vector of dimension $L$ is firstly mapped to a vector of dimension $W$ via a fully connected (FC) layer and then projected back to the dimension $L$ via the other FC layer. The output vector is then added to the input vector forming a residue connection. The hyperparameter width $W$ controls the expressiveness of the NQS ansatz.}
\label{fig:vmc_ansatz}
\end{figure}

{\bf Neural quantum states.}
For NQS, the probability amplitude is given by the output of neural networks with the input the computational basis, i.e.
\begin{equation}
 \langle i_1i_2\cdots i_L \vert \psi\rangle = f(i_1i_2\cdots i_L),
\end{equation}
where $f$ corresponds to the neural network. In this work, we employ a deep neural network with the architecture shown in Fig. \ref{fig:vmc_ansatz}. Given the small system size, we do not run the variational Monte Carlo method with sampling to optimize the NQS. Instead, similar to the MPS and PEPS cases, we directly evaluate the neural network output on all $2^L$ computational basis inputs to first obtain the (unnormalized) full wavefunction and optimize infidelity or energy based on the wavefunction. The total number of trainable parameters is of order $O(LWD)$, where $D$ is the number of repetitions of the ResBlock and $W$ is the width of intermediate fully connected layers. The network width $W$ controls the expressiveness of the NQS ansatz.

\begin{figure}[t]\centering
	\includegraphics[width=0.48\textwidth]{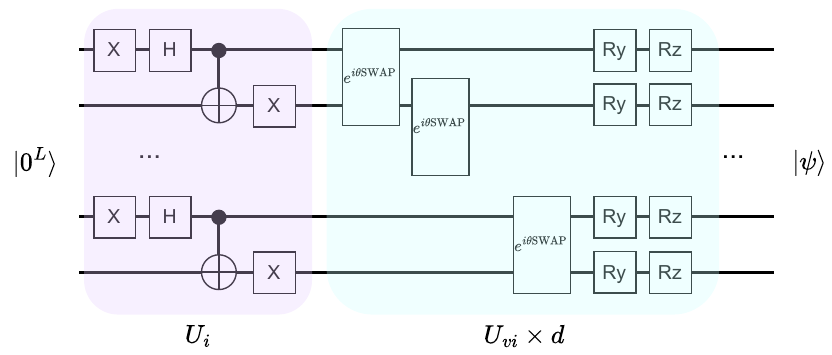}
	\caption{\textbf{VQE ansatz employed in this work.} For 1D lattice, a parameterized swap gate on qubits $1$ and $L$ is omitted. For 2D lattices, these parameterized swap gates are aligned with lattice bonds between the nearest neighbor sites. The initialization block prepares the initial states as copies of Bell pairs $1/\sqrt{2}(\vert 01\rangle-\vert 10\rangle)$. The variational blocks are repeated for $d$ times with different parameters. Ry and Rz are notations for $e^{-i\theta Y}$ and $e^{-i \theta Z}$ gates, respectively. The output state from the PQC is employed for optimization and approximating target states.  }
\label{fig:vqe_ansatz}
\end{figure}

{\bf Variational quantum eigensolver states.} In VQE methods, we use the output quantum states from the parameterized quantum circuits (PQC) and the trainable parameters correspond to rotation angles of quantum gates on the PQC. The PQC $U=U_v U_i$ consists of an initialization block $U_i$ and $d$ variational blocks $U_v=\prod_i^d U_{vi}$. The initialization block prepares the Bell pair $1/\sqrt{2}(\vert 01\rangle-\vert 10\rangle)$ on nearest neighbor qubits so that the initial state is in the $J_{tot}=0$ sector. Each variation block is given by: $U_{vi}=\prod_j^L e^{-i\theta^{(0)}_{j} Z_j}e^{-i \theta^{(1)}_{j} Y_j}\prod_{\langle jk\rangle} e^{-i \theta^{(2)}_{jk} \text{SWAP}_{jk}}$, where the total number of circuit parameters $\theta$ is of the order $O(Ld)$ and $\text{SWAP}_{jk}=1/2(X_jX_k+Y_jY_k+Z_jZ_k+I_jI_k)$ is the swap gate on qubits $j$ and $k$. These parameterized swap gates are applied on the bond between nearest neighbor sites according to the lattice geometry. For 2D square lattice, these two-qubit gates are first applied on all rows following all columns without periodic bonds across the opposite boundaries. For 1D lattice, these two-qubit gates are applied on the system in a ladder fashion with a final periodic bond gate, i.e. the entangling gates apply on the qubit pairs  $(1, 2), (2, 3), \cdots (L-1, L), (L, 1)$.  A schematic of the PQC structure investigated in this work is shown in Fig. \ref{fig:vqe_ansatz}. The approximate state we use is $\vert \psi\rangle =U\vert 0^L\rangle$ accordingly. Our numerical simulation for expectation value and fidelity omits quantum noise error and shot sampling error inevitable on real quantum devices. The circuit parameter gradients are obtained via automatic differentiation in our simulation and can be obtained via parameter shift \cite{Crooks2019} on quantum processors.

{\bf Vector states.} Since the system size under investigation is small, we can directly build up the so-called vector states ansatz where the wavefunction amplitude of each basis is directly modeled as a trainable parameter. For a spin-1/2 system of size $L$, $2^L$ complex parameters are required, namely, we treat $\langle i_1i_2\cdots i_L \vert \psi\rangle$ as variational parameters. In our work, the real part and imaginary part of these complex parameters are independently initialized (Gaussian distribution of mean $0$ and standard deviation $0.1$) and optimized. This is because the Adam optimizer does not work as expected with complex training parameters.
This ansatz is expected to encode the least physical prior and be the most unbiased.

~\\

\noindent\textbf{Acknowledgments} We thank Yi-Fan Jiang, Zi-Xiang Li, and Lei Wang for their valuable discussions. SXZ acknowledges support from a startup grant at CAS-IOP.

%


\onecolumngrid

\clearpage
\newpage
\widetext

\begin{center}
\textbf{\large Supplemental Material for ``Effective Temperature in Approximate Quantum Many-body States''}
\end{center}

\renewcommand{\thefigure}{S\arabic{figure}}
\setcounter{figure}{0}
\renewcommand{\theequation}{S\arabic{equation}}
\setcounter{equation}{0}
\renewcommand{\thesection}{\Roman{section}}
\setcounter{section}{0}
\setcounter{secnumdepth}{4}


\section{Training dynamics for effective temperature toward the ground state}

\begin{figure}[t]\centering
	\includegraphics[width=0.5\textwidth]{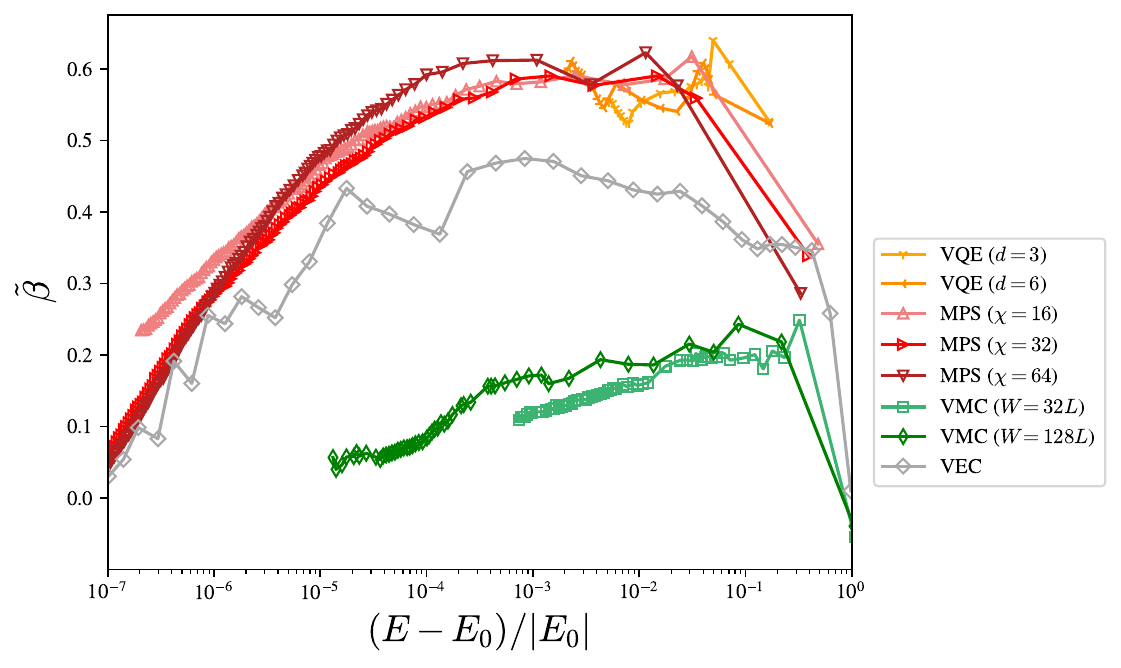}
	\caption{Training dynamics for $\tilde{\beta}$ for 1D XXZ model $L=12$ with $J_x=J_y=1, J_z=0.8, h_z=0.02$ and energy objective.}
\label{fig:1d_energy}
\end{figure}

\begin{figure}[t]\centering
	\includegraphics[width=0.5\textwidth]{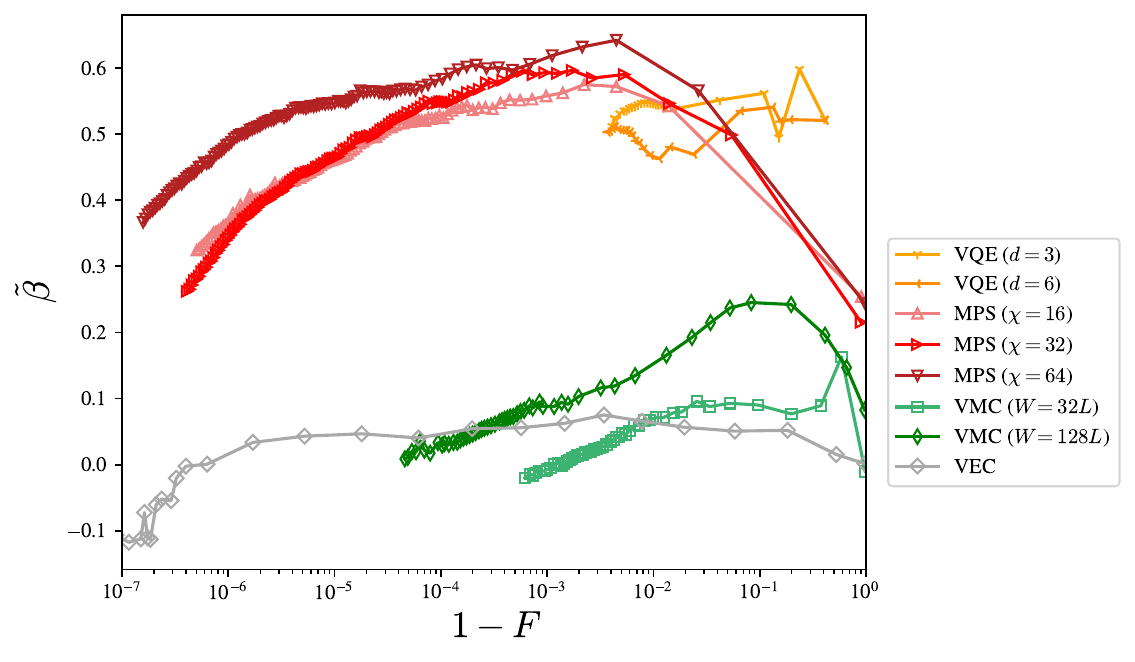}
	\caption{Training dynamics for $\tilde{\beta}$ for 1D XXZ model $L=12$ with $J_x=J_y=1, J_z=0.8, h_z=0.02$ and ground state fidelity objective.}
\label{fig:1d_fidelity}
\end{figure}

\begin{figure}[t]\centering
	\includegraphics[width=0.55\textwidth]{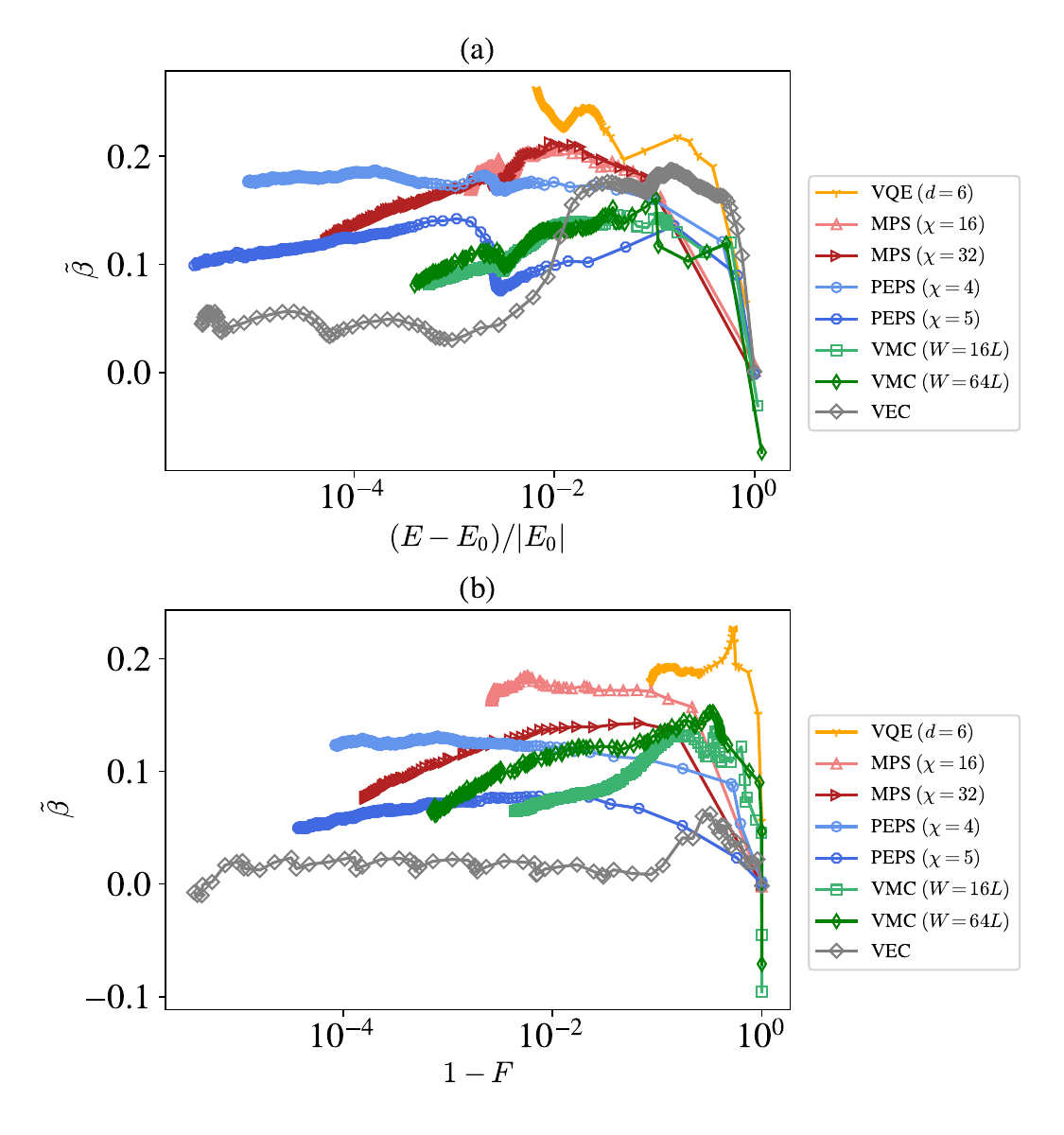}
	\caption{Training dynamics for $\tilde{\beta}$ for 2D XXZ model with $J_x=J_y=1, J_z=1.5$ on $4\times 4$ lattice with energy (a) and ground state fidelity objective (b).}
\label{fig:44_1_1_1.5_beta}
\end{figure}

\begin{figure}[t]\centering
	\includegraphics[width=0.55\textwidth]{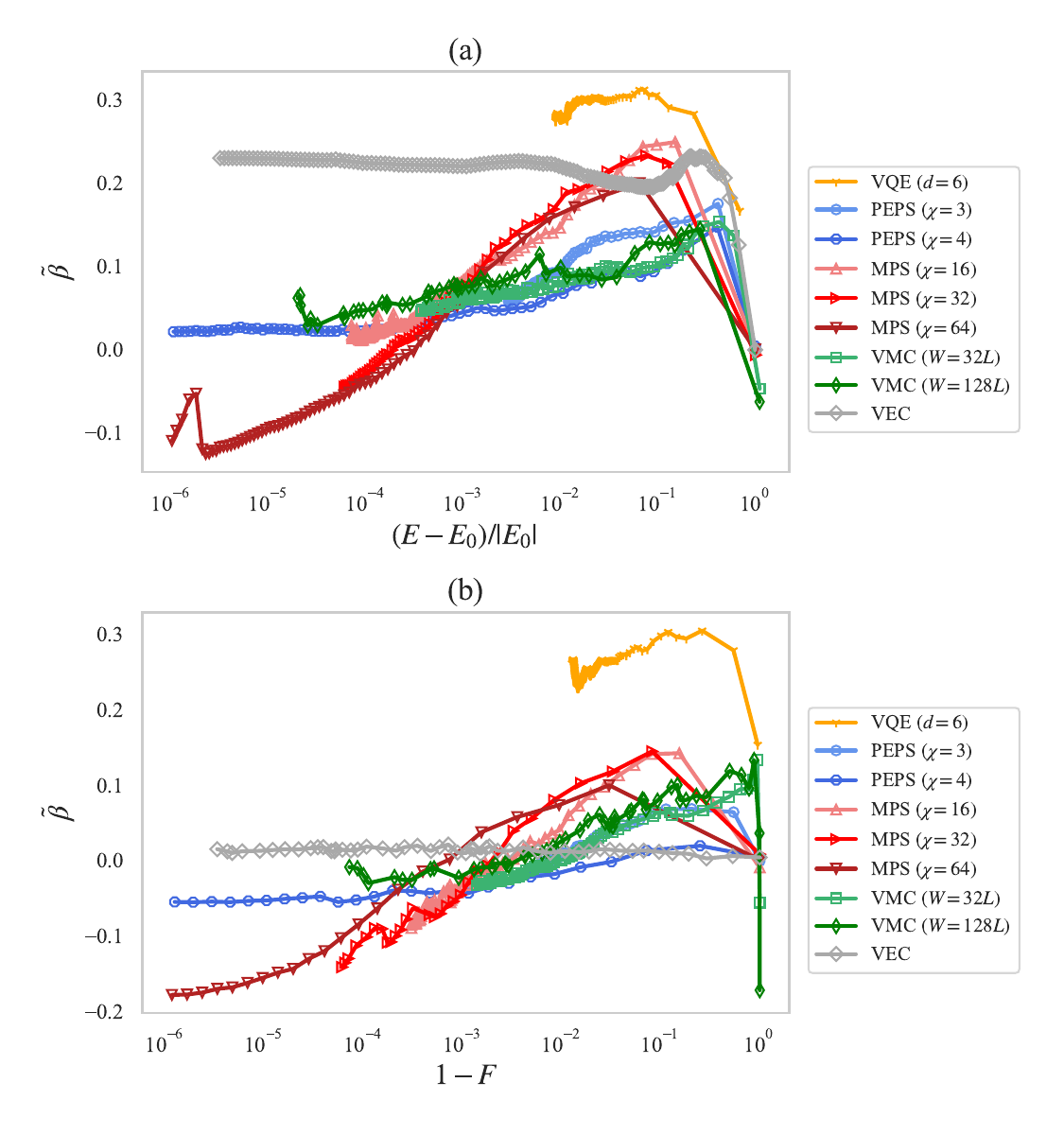}
	\caption{Training dynamics for $\tilde{\beta}$ for 2D XXZ model with $J_x=J_y=1, J_z=0.8$ on $4\times 3$ lattice with energy (a) and ground state fidelity objective (b).}
\label{fig:4_3_1_1_0.8_beta}
\end{figure}

We conduct numerical experiments on different models and find that the dynamics for $\tilde{\beta}$ show a universal pattern. Fig. \ref{fig:1d_energy} and Fig. \ref{fig:1d_fidelity} show the training dynamics for the 1D XXZ model with $\mathcal{L}_E$ and $\mathcal{L}_F$ as objectives, respectively. The system Hamiltonian for the 1D XXZ model with periodic boundary conditions is defined as:
\begin{equation}
    H = \sum_i J_x X_i X_{i+1} + J_y Y_i Y_{i+1} + J_z Z_i Z_{i+1} + h_i Z_i.
\end{equation}
Similarly, the training dynamics for 2D XXZ models in antiferromagnetic phases is shown in Fig. \ref{fig:44_1_1_1.5_beta} which complements the results in the main text where the target ground state is in spin-flipping phases. Besides, we also evaluated $\tilde{\beta}$ with different 2D system size $4\times 3$, as shown in Fig. \ref{fig:4_3_1_1_0.8_beta}.

From these training dynamics results toward ground states on different dimensions and different phases, several important features remain consistent as universal patterns for approximate ground states.
\begin{enumerate}
\item Optimized approximate ground states admit very small $\tilde{\beta}$s indicating a nearly flat spectrum across excited states. 
\item During training, $\tilde{\beta}$ first increases and then decreases in general, implying an upper bound for the possible $\tilde{\beta}$ of a given ansatz. Moreover, a smaller $\tilde{\beta}$ often indicates a better accuracy at the late stage of training. Besides, initial $\tilde{\beta}\approx 0$ unless the ansatz has a strong physical prior. For example, in the VQE case, the PQC structure and the initialization render the state very close to the valence bond state at the beginning, resulting in a higher decay factor $\tilde{\beta}$. 
\item For each method, the training dynamics is qualitatively similar for both energy and fidelity objectives though $\tilde{\beta}$ values are typically smaller when optimizing fidelity. The energy objective imposes unequal penalties for different excited state components, i.e. the higher energy the excited state has, the larger penalty it receives. On the contrary, the fidelity objective equally penalizes all excited states which could result in smaller or even negative $\tilde{\beta}$.  Since the fidelity objective shows no preference for different excited states, the finite $\tilde{\beta}$ value in this case can be attributed to the physical prior encoded in the ansatzes. Indeed, for the most unbiased ansatz VEC, $\tilde{\beta}$ is very close to 0 when optimizing $\mathcal{L}_f$. In summary, both the objective structures and the ansatz structures determine the effective temperature of resultant approximate ground states.
\item The same family of ansatzes with different hyperparameters, eg. VMC of different widths or tensor networks with different bond dimensions, usually display similar training dynamics. This observation further supports the conclusion that effective temperature can reflect the intrinsic properties of different numerical methods. 
\end{enumerate}

\begin{figure}[t]\centering
	\includegraphics[width=0.5\textwidth]{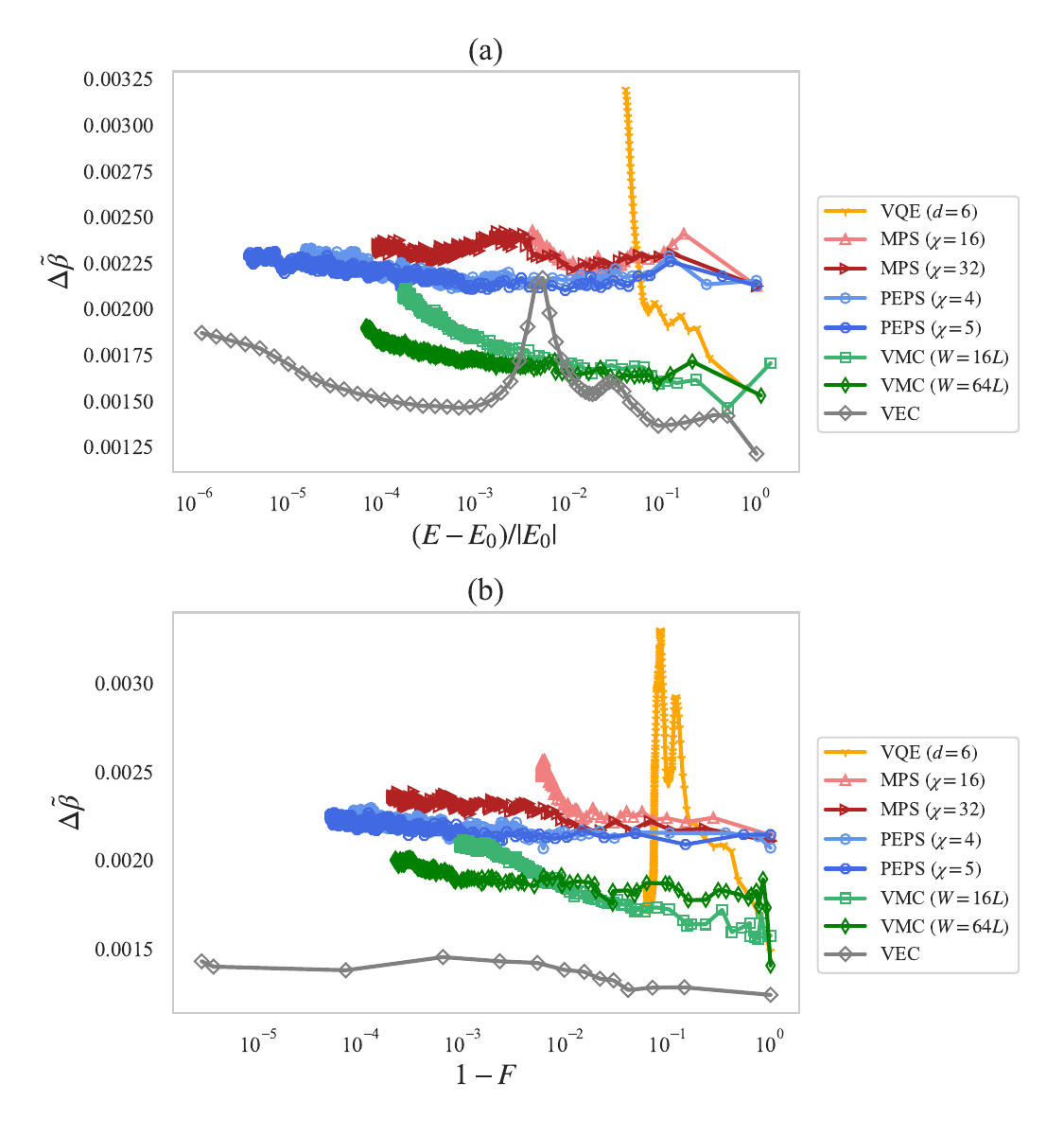}
	\caption{Training dynamics for the uncertainty $\Delta \tilde{\beta}$ of fitted $\tilde{\beta}$ for 2D XXZ model with $J_x=J_y=1, J_z=0.8$ on $4\times 4$ lattice with energy (a) and ground state fidelity objective (b).}
\label{fig:4_4_1_1_0.8_delta}
\end{figure}

In addition, we also show the error bar $\Delta \tilde{\beta}$ given by the exponential fitting in Fig. \ref{fig:4_4_1_1_0.8_delta}. The uncertainty of the predicted decay factor is relatively small compared to the $\tilde{\beta}$ value, implying that the exponential scaling is a very good description for the spectrum decomposition.

\section{Training dynamics for the scaling factor toward the ground state}

\begin{figure}[t]\centering
	\includegraphics[width=0.5\textwidth]{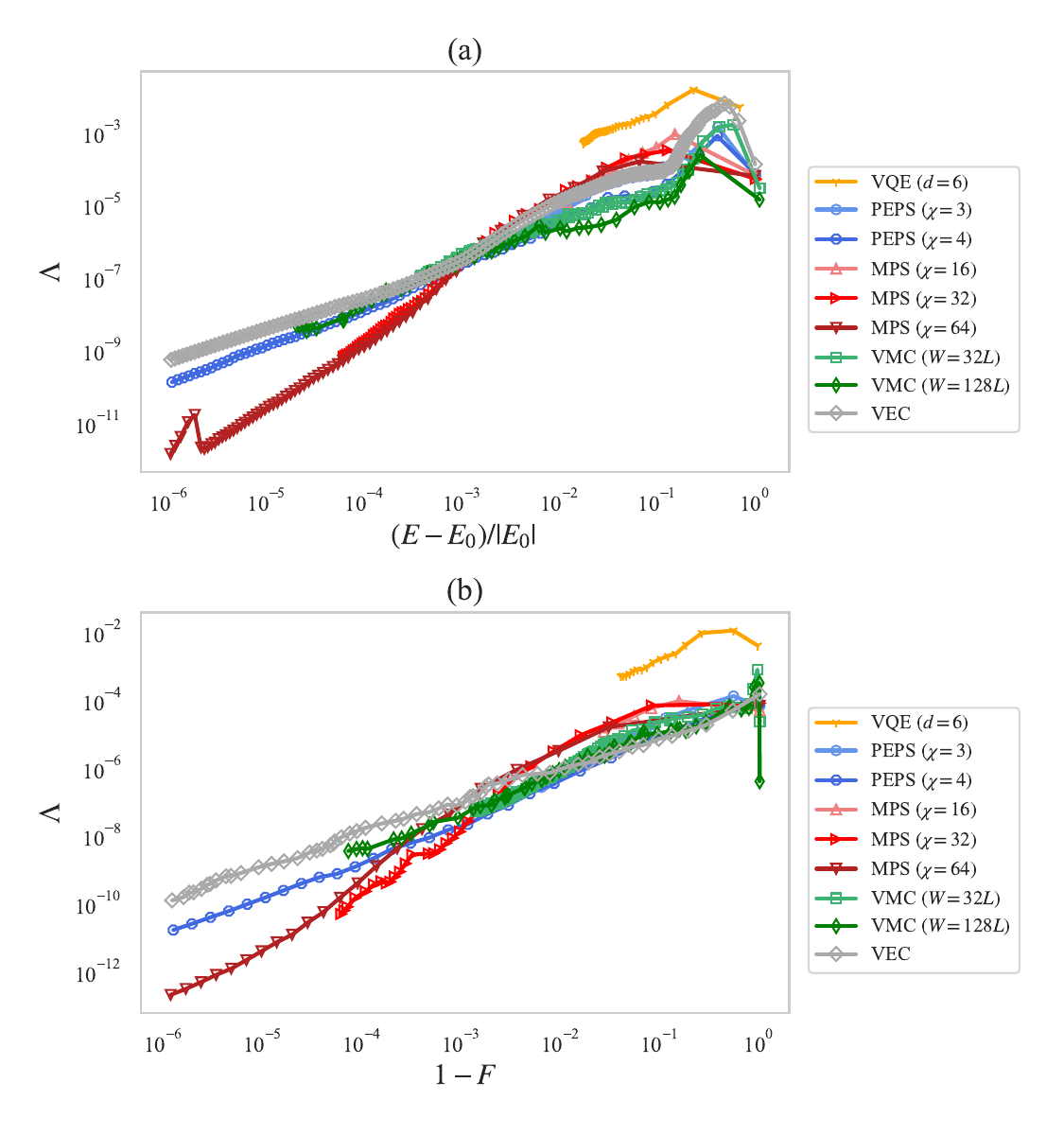}
	\caption{Training dynamics for $\Lambda$ for 2D XXZ model with $J_x=J_y=1, J_z=0.8$ on $4\times 3$ lattice with energy (a) and ground state fidelity objective (b).}
\label{fig:4_3_1_1_0.8_lambda}
\end{figure}

\begin{figure}[t]\centering
	\includegraphics[width=0.5\textwidth]{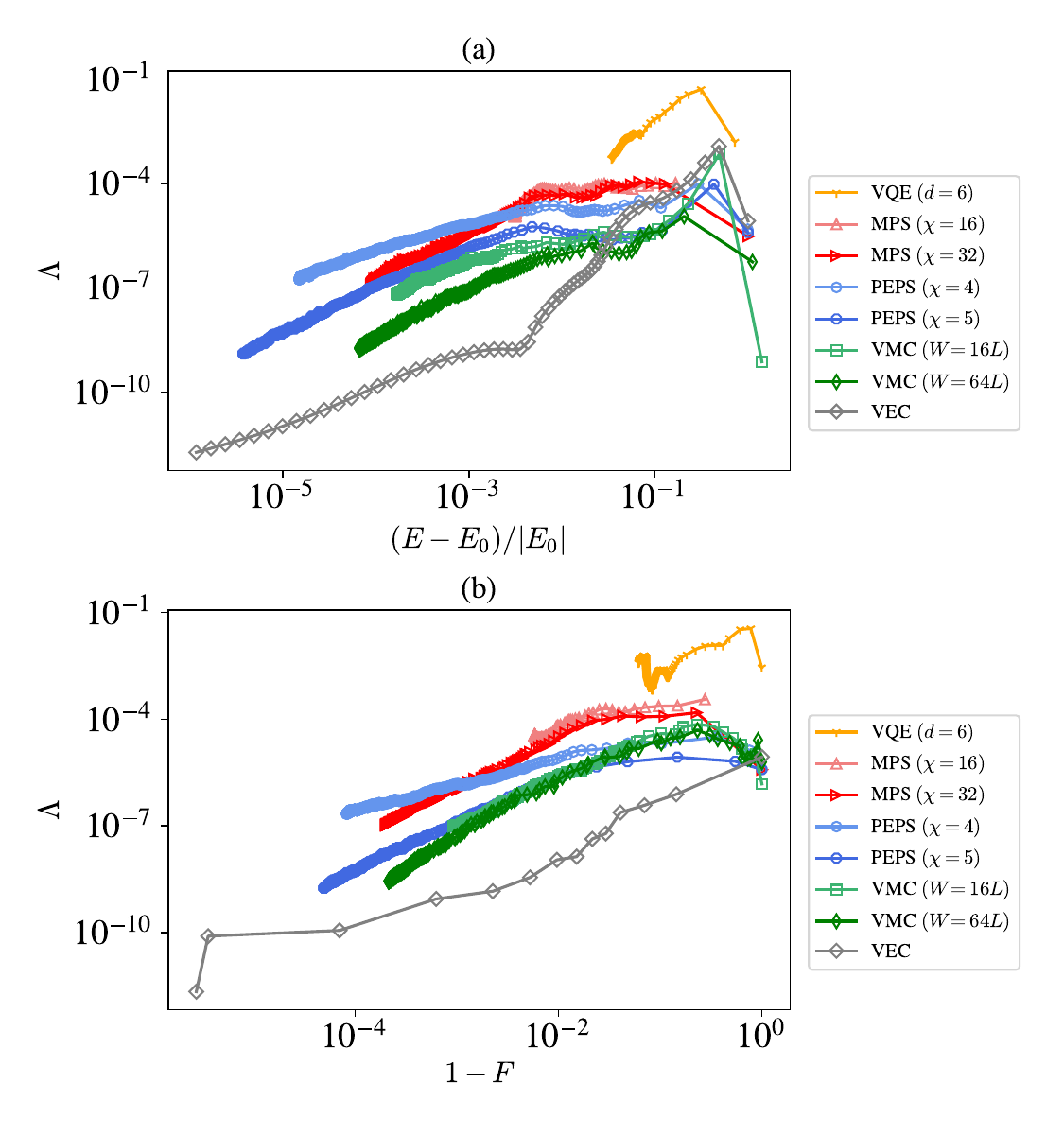}
	\caption{Training dynamics for $\Lambda$ for 2D XXZ model with $J_x=J_y=1, J_z=0.8$ on $4\times 4$ lattice with energy (a) and ground state fidelity objective (b).}
\label{fig:4_4_1_1_0.8_lambda}
\end{figure}

\begin{figure}[t]\centering
	\includegraphics[width=0.5\textwidth]{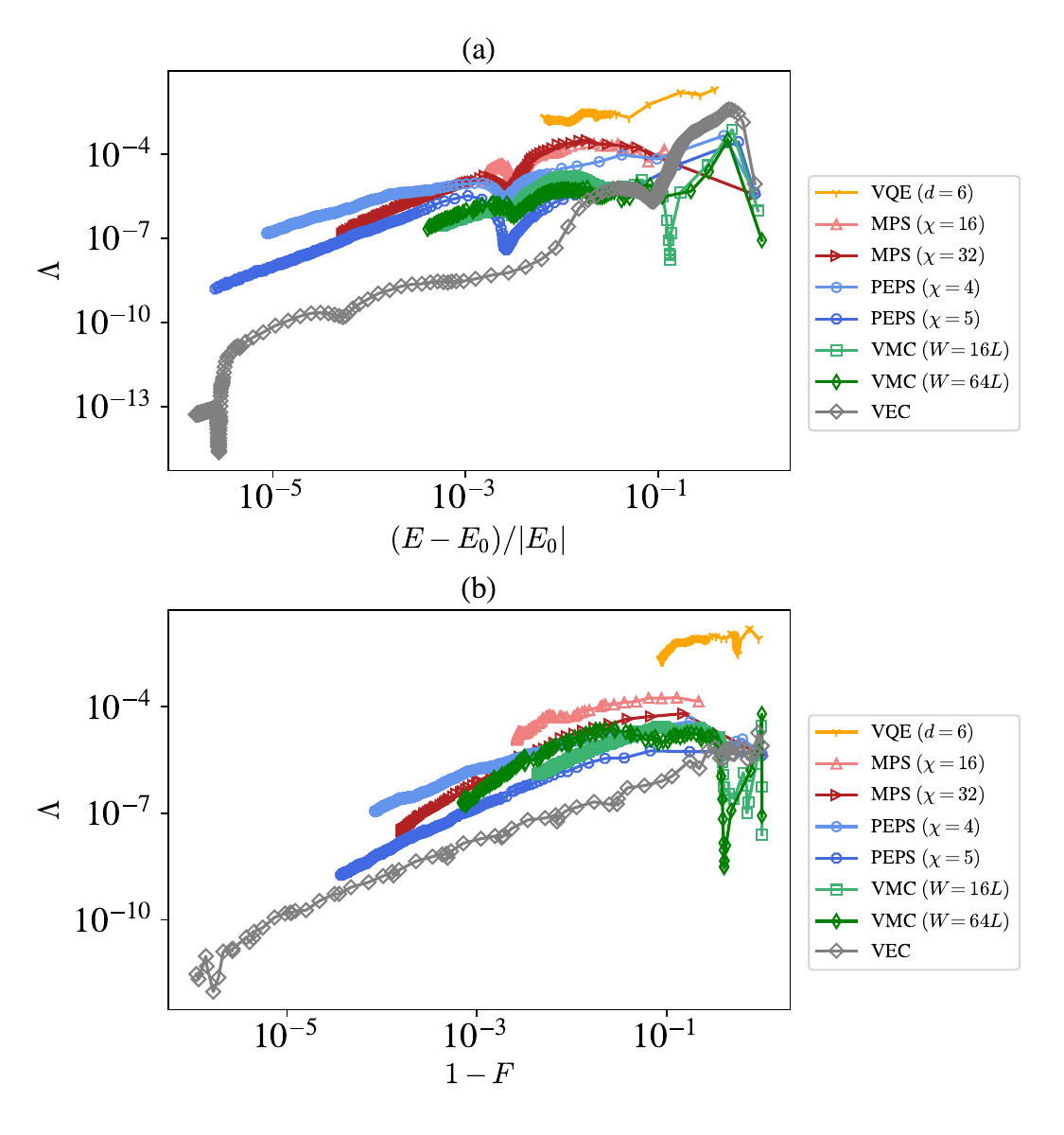}
	\caption{Training dynamics for $\Lambda$ for 2D XXZ model with $J_x=J_y=1, J_z=1.5$ on $4\times 4$ lattice with energy (a) and ground state fidelity objective (b).}
\label{fig:4_4_1_1_1.5_lambda}
\end{figure}

When fitting the exponential decay spectrum, two parameters are required in the form $\Lambda e^{-\tilde{\beta}\varepsilon}$. We focus on the training dynamics for $\Lambda$ in this section. Parameter $\Lambda$ is more directly related to the overall approximation accuracy, as the infidelity can be roughly estimated as $1-F = \int_{\varepsilon=E_{min}}^{\varepsilon=E_{max}} d\varepsilon\rho(\varepsilon) \Lambda e^{-\tilde{\beta} \varepsilon}\propto \Lambda$, where $\rho(\varepsilon)$ is the density of states for the system. Some typical training dynamics for $\Lambda$ toward ground state are shown in Fig. \ref{fig:4_3_1_1_0.8_lambda}, \ref{fig:4_4_1_1_0.8_lambda}, and \ref{fig:4_4_1_1_1.5_lambda}. We can observe an evident linear correlation during training between $\Lambda$ and $1-F$.

\section{Training dynamics for effective temperature toward imaginary-time evolved states}

\begin{figure}[t]\centering
	\includegraphics[width=0.8\textwidth]{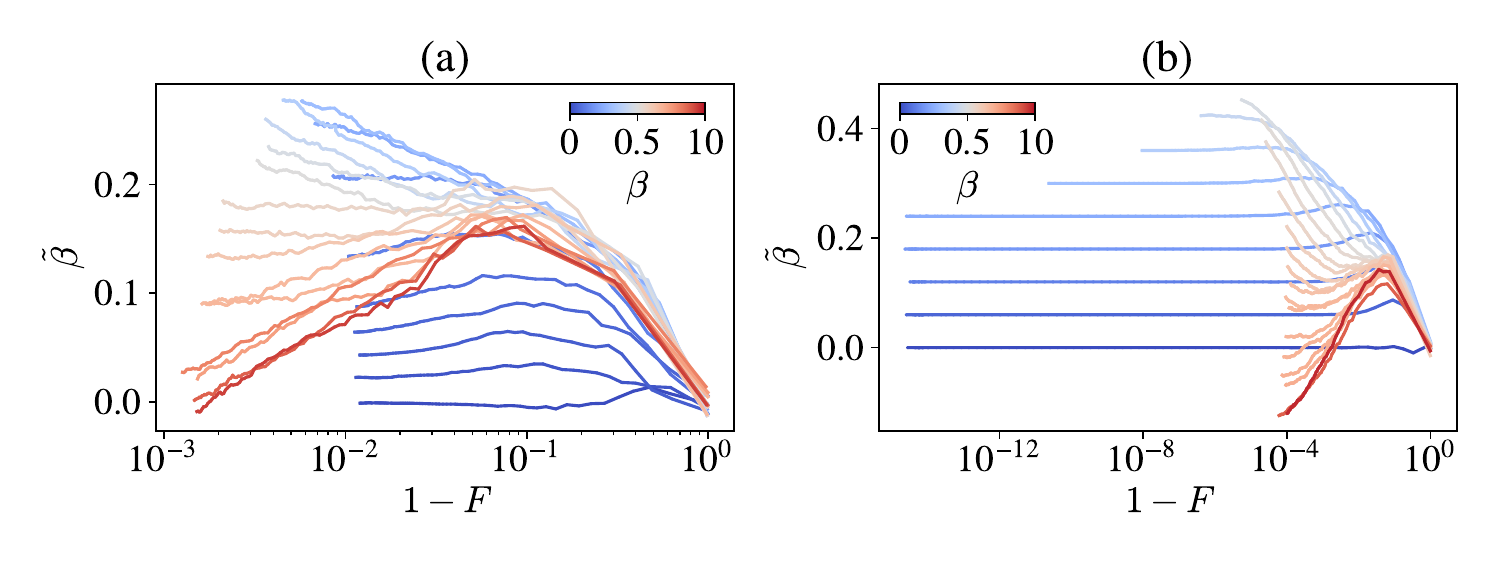}
	\caption{Training dynamics for $\tilde{\beta}$ for 2D XXZ model with $J_x=J_y=1, J_z=0.8$ on $4\times 3$ lattice and ITES $\beta$ targets. MPS ansatzes with bond dimensions $\chi=16$ (a) and $\chi=32$ (b) are employed. }
\label{fig:ites_dynamics}
\end{figure}

We report training dynamics toward ITES in this section, and the limit $\beta\rightarrow \infty$ reduces to the results for approximate ground states. A typical instance of such training dynamics with varying beta is shown in Fig. \ref{fig:ites_dynamics}. For optimization targeting high temperature ITES (low $\beta$), $\tilde{\beta}$ first increases and then saturates to the target $\beta$ value. Conversely, for low temperature (high $\beta$) ITES targets, $\tilde{\beta}$ first increases and then decreases during training and the value of $\tilde{\beta}$ that can be reached is upper bounded by $\beta^*$. The distinct training dynamical behaviors are also a reflection of the two-stage behavior. The training dynamics in low temperature regime is consistent with the one in approximating ground states. Via ITES of different temperatures, the results for ground state simulation can be taken as an interpolation from ITES and the maximal $\tilde{\beta}$ reached during training toward ground state is also connected to $\beta^*$ in ITES training.

\section{Spectrum pattern for approximate states from different ansatzes}

\begin{figure}[t]\centering
	\includegraphics[width=0.7\textwidth]{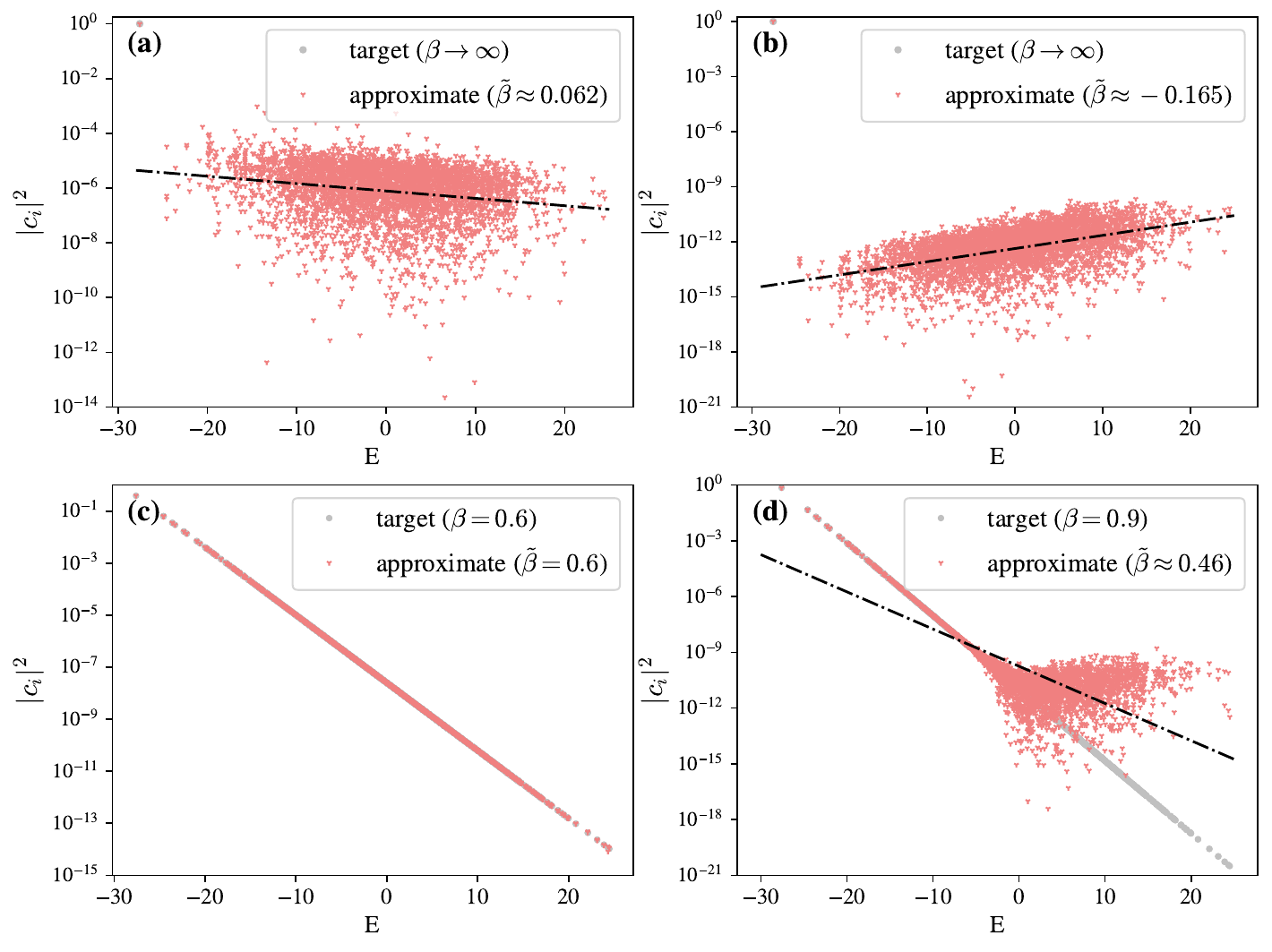}
	\caption{2D XXZ model on $4\times 3$ square lattice with $J_x=J_y=1, J_z=0.8$. MPS ansatz with bond dimension $\chi=64$ is employed. The optimization objective is the infidelity between approximate states and the target states. The target states are chosen as the ground states for (a)(b) and imaginary-time evolved states $\vert \phi(\beta)\rangle$ with $\beta=0.6$ (c) and $\beta=0.9$ (d). The overlaps with different energy eigenstates are depicted as scatter plots in red and gray for approximate states and target states, respectively. The optimization is at the intermediate stage for (a) (less trained)  while the optimization is converged for (b) (well trained) toward the ground state. The logarithmic overlaps with excited eigenstates show a linear pattern during training with varying fitted slope $\tilde{\beta}$, aka. inverse effective temperature. The fitted line is shown in dashdot in the figure.  When the target state is imaginary-time evolved state, there is a phase transition in the spectrum behaviors of the approximate state. For small $\beta$ in (c), the optimized approximate state exhibits a near-perfect correspondence with the target state overlap, with a fitted slope $\tilde{\beta}$ matching $\beta = 0.6$. Conversely, for large $\beta$ in (d), the overlaps from approximate states exhibit a distinct behavior -- an exponential decay in the lower energy regime and a plateau in the higher energy regime. This behavior leads to a poor linear fit, characterized by a deviating slope $\tilde{\beta}<\beta$. The transition is not induced by numerical machine precision issue since the alignment is perfect for $\vert c_i\vert^2\sim 10^{-15}$ in (c) while the plateau is around $10^{-10}$ in (d), much higher than the machine precision floor.}
\label{fig:mps64_overlap}
\end{figure}

\begin{figure}[t]\centering
	\includegraphics[width=0.7\textwidth]{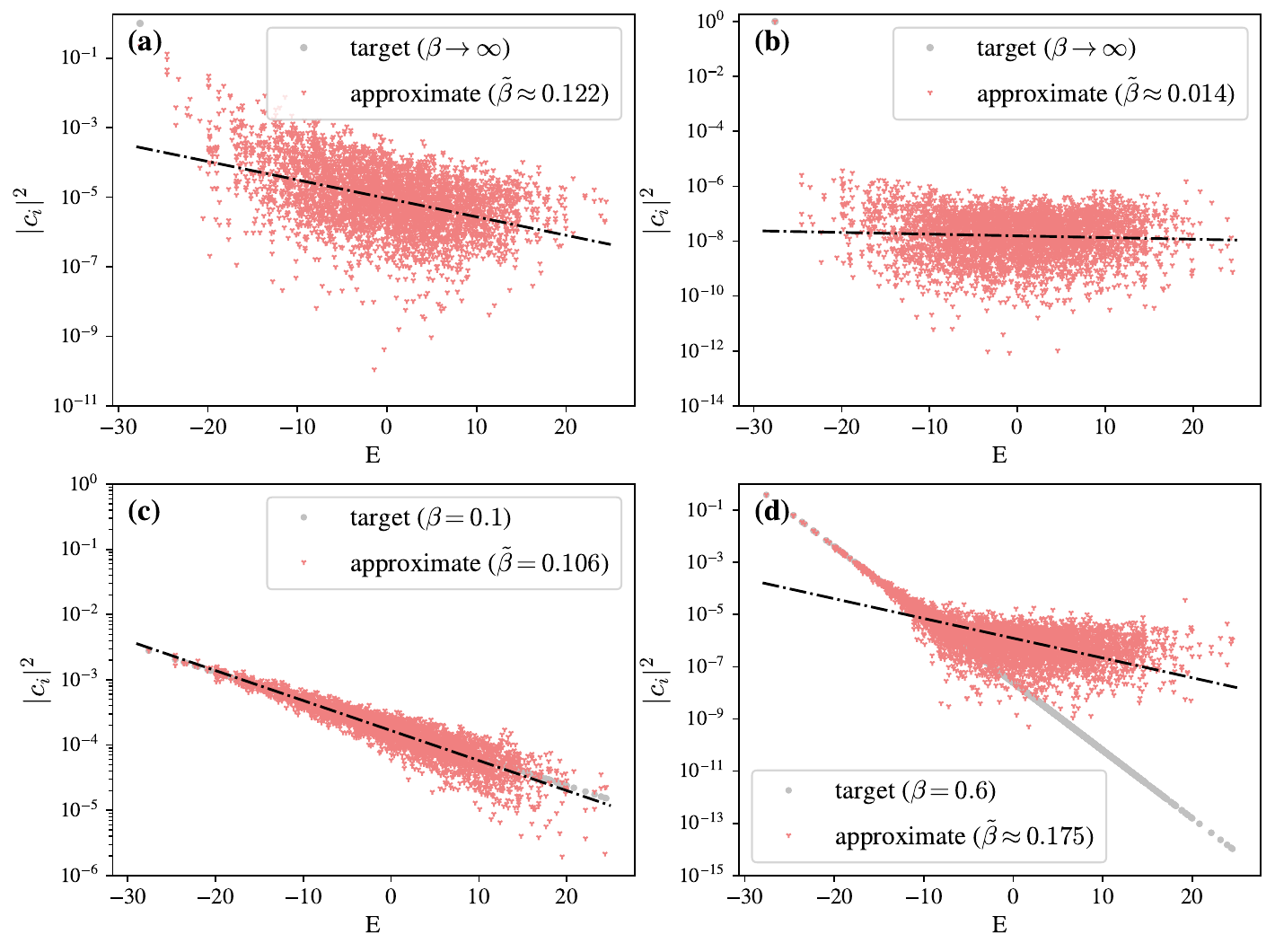}
	\caption{2D XXZ model on $4\times 3$ square lattice with $J_x=J_y=1, J_z=0.8$. VMC ansatz with neural network width $W=128L$ is employed. The optimization objective is the infidelity between approximate states and the target states. The target states are chosen as the ground states for (a)(b) and imaginary-time evolved states $\vert \phi(\beta)\rangle$ with $\beta=0.1$ (c) and $\beta=0.6$ (d). The overlaps with different energy eigenstates are depicted as scatter plots in red and gray for approximate states and target states, respectively. The optimization is at the intermediate stage for (a) (less trained)  while the optimization is converged for (b)(c)(d). The logarithmic overlaps with excited eigenstates show a linear pattern during training with varying fitted slope $\tilde{\beta}$, aka. inverse effective temperature. The fitted line is shown in dashdot line in the figure.  When the target state is imaginary-time evolved state, there is a phase transition in the spectrum behaviors of the approximate state. For small $\beta$ in (c), the optimized approximate state exhibits a near-perfect correspondence with the target state overlap, with a fitted slope $\tilde{\beta}$ matching $\beta = 0.6$. Conversely, for large $\beta$ in (d), the overlaps from approximate states exhibit a distinct behavior -- an exponential decay in the lower energy regime and a plateau in the higher energy regime. This behavior leads to a poor linear fit, characterized by a deviating slope $\tilde{\beta}<\beta$.}
\label{fig:vmc_overlap}
\end{figure}

\begin{figure}[t]\centering
	\includegraphics[width=0.7\textwidth]{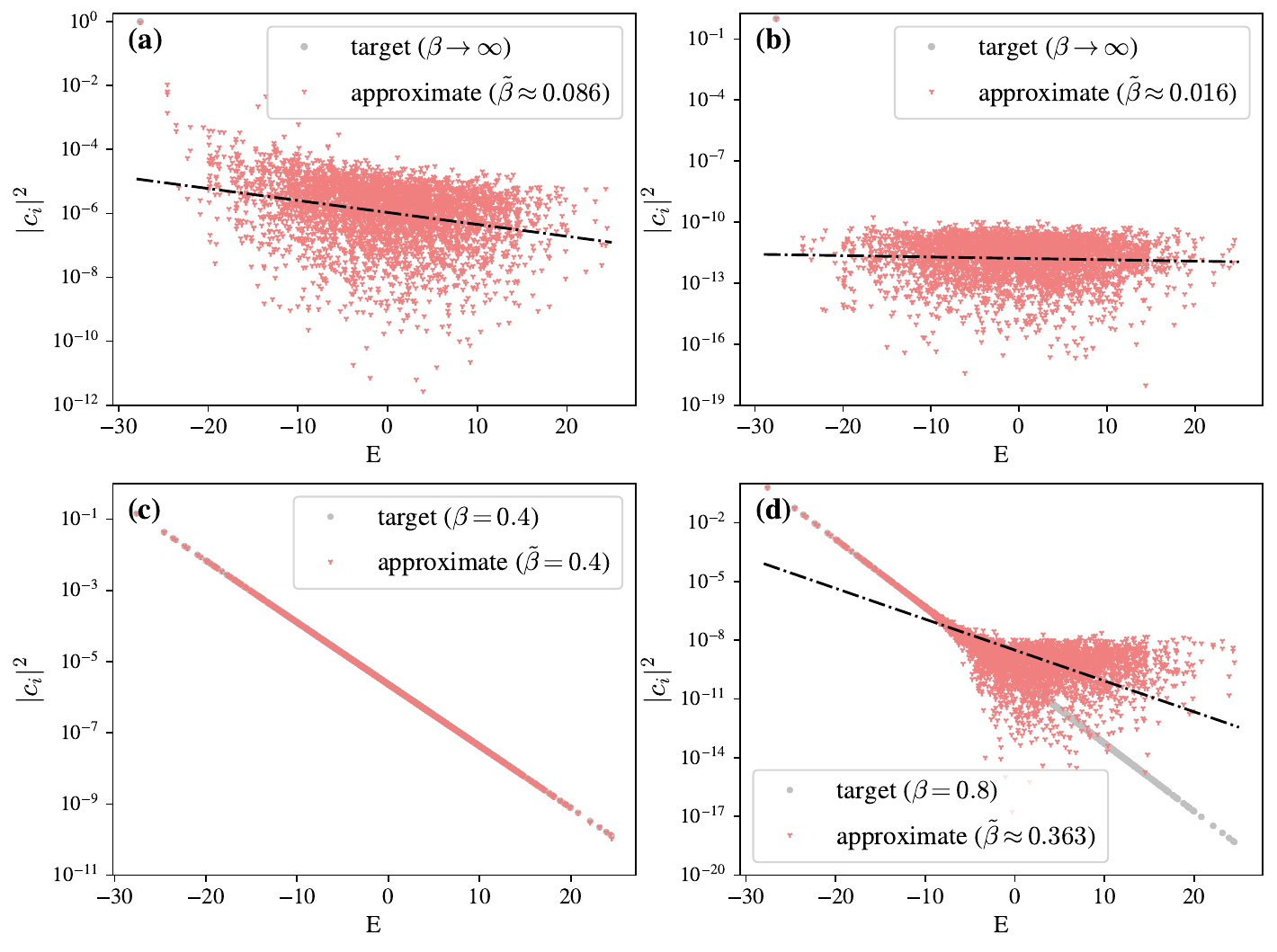}
	\caption{2D XXZ model on $4\times 3$ square lattice with $J_x=J_y=1, J_z=0.8$. PEPS ansatz with bond dimension $\chi=4$ is employed. The optimization objective is the energy expectation for (a)(b) where the target states are the ground states with minimal energy. The training objective is the infidelity against imaginary-time evolved states $\vert \phi(\beta)\rangle$ with $\beta=0.4$ (c) and $\beta=0.8$ (d). The overlaps with different energy eigenstates are depicted as scatter plots in red and gray for approximate states and target states, respectively. The optimization is at the intermediate stage for (a) (less trained)  while the optimization is converged for (b)(c)(d). The logarithmic overlaps with excited eigenstates show a linear pattern during training with varying fitted slope $\tilde{\beta}$, aka. inverse effective temperature. The fitted line is shown in dashdot in the figure.  When the target state is imaginary-time evolved state, there is a phase transition in the spectrum behaviors of the approximate state. For small $\beta$ in (c), the optimized approximate state exhibits a near-perfect correspondence with the target state overlap, with a fitted slope $\tilde{\beta}$ matching $\beta$. Conversely, for large $\beta$ in (d), the overlaps from approximate states exhibit a two-stage behavior -- an exponential decay in the lower energy regime and a plateau in the higher energy regime. This distinct behavior leads to a poor linear fit, characterized by a deviating slope $\tilde{\beta}<\beta$.}
\label{fig:peps4_overlap}
\end{figure}

In this section, we show more scatter plots for $(\varepsilon_i, |c_i|^2)$ pairs from approximate states based on different forms of ansatzes. Fig. \ref{fig:mps64_overlap}, \ref{fig:vmc_overlap} and \ref{fig:peps4_overlap} show the spectral decomposition of approximate states generated by MPS, VMC, and PEPS ansatzes, respectively. The approximate ground states in the former two cases are obtained via minimizing infidelity objective while for the latter, energy objective is employed.  For ground state targets, the fitted factor $\tilde{\beta}$ continuously varies with the optimization progress. For ITES targets, the spectrum from each type of approximate states exhibits the two-stage behavior with different critical temperatures $\beta^*$.

We emphasize that the effective temperature is meaningful and argue against the misconception that the spectral plateau is merely an artifact of numerical precision. Our reasoning is as follows: 
\begin{enumerate}
\item Many flat spectrum plateaus give $|c_i|^2$ in the order $10^{-4}\sim 10^{-8}$, which are numerically significant and much higher than the machine precision floor $10^{-16}$ for double precision (complex128) arithmetic. 

\item Many approximate ITES instances exhibit the two-stage behavior such as Fig. \ref{fig:mps64_overlap}, demonstrate distinct overlap behaviors for the same energy levels across different target $\beta$ values. For example, points with target overlap $|c_i|^2=10^{-10}$ strictly follow the exponential decay in approximate high-temperature ITES while the overlap point of the same order fails to do so by forming a plateau for approximate low-temperature ITES. In other words, the plateau formed in the higher energy regime for approximate low-temperature ITES is not due to the na\"ive interpretation that the magnitudes of overlaps go below the machine precision floor.

\item The critical effective temperature $\beta^*$s differ among different ansatzes. This fact further implies that the effective temperature is numerically meaningful. If the underlying mechanism were solely the machine precision floor, $\beta^*$ would be the same across different methods.
\end{enumerate}


\begin{figure}[t]\centering
	\includegraphics[width=0.7\textwidth]{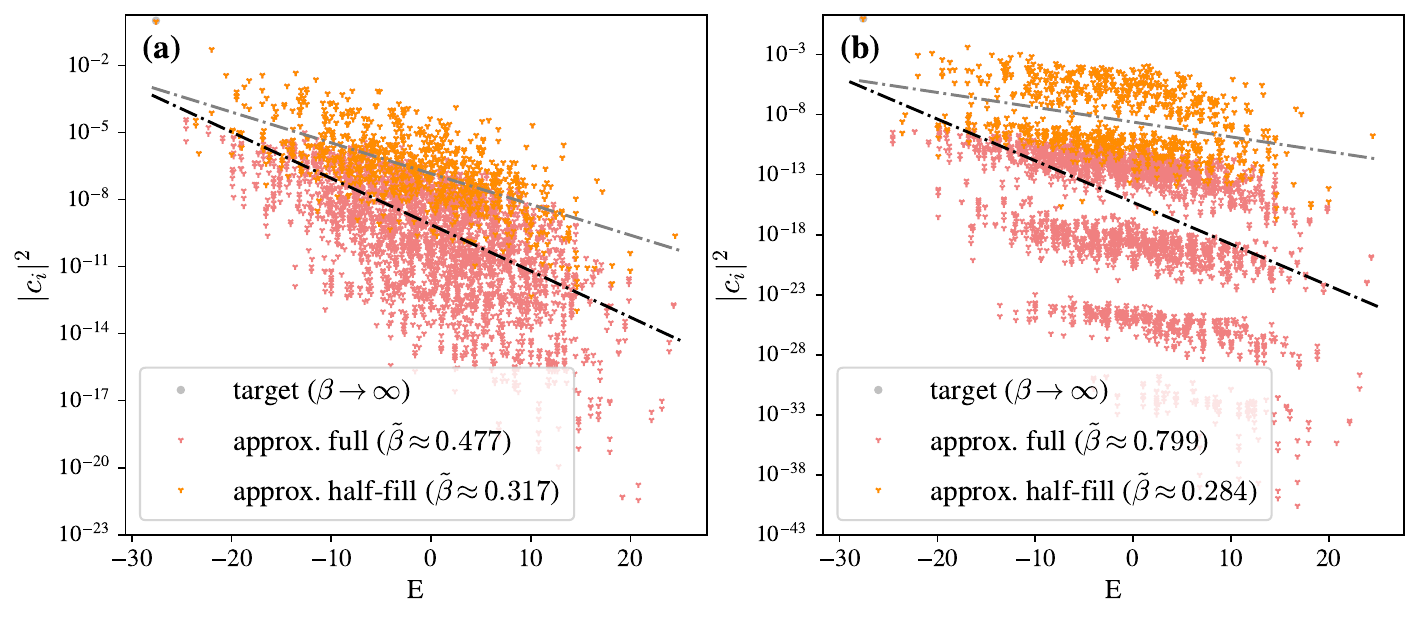}
	\caption{2D XXZ model on $4\times 3$ square lattice with $J_x=J_y=1, J_z=0.8$. VQE ansatz with depth $d=6$ is employed. The optimization objective is the energy expectation targeting the ground state. The overlaps with different energy eigenstates are depicted as scatter plots in red and gray for optimized approximate states and target states, respectively. Specifically, we highlight the overlap with eigenstates in the half-filling sector with orange color. As we can see, the overlap points become separated for different charge sectors during VQE training which is a unique feature absent in other ansatzes. The optimization is at the intermediate stage for (a) (less trained)  while the optimization is converged for (b) (well trained). The overlaps with excited eigenstates show a linear pattern during training with varying fitted slope $\tilde{\beta}$, aka. inverse effective temperature. The fitted line for all overlap points is shown in dashdot black line while the fitted line for overlap points within half-filling sector is shown in dashdot gray line. The half-filling sector spectrum contribution is dominant since the exact ground state lives in this sector.}
\label{fig:vqe6_overlap}
\end{figure}

The spectral decomposition for VQE states is special due to the specific structure and state preparation protocol we used in this work. The corresponding scatter plot at the early and late stages of ground state training is shown in Fig. \ref{fig:vqe6_overlap}. As we can see, several different branches emerge in the spectral decomposition with more training steps. These branches are labeled by different charge numbers as the system Hamiltonian is $U(1)$ symmetric. Therefore, the decay factor by considering spectrum contribution from all charge sectors is larger than the one fitted only in the half-filling sector which is highlighted in orange in the figure. We always present the decay factor extracted from the half-filling sector in VQE cases in this work.

\begin{figure}[t]\centering
	\includegraphics[width=0.8\textwidth]{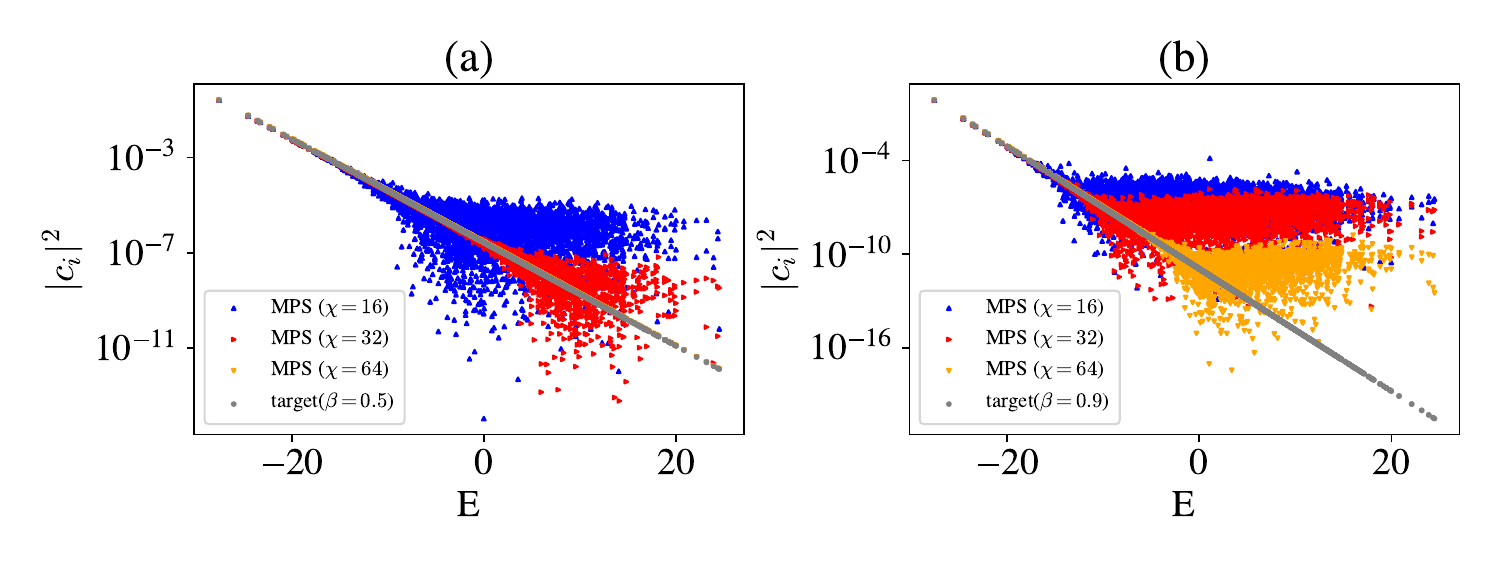}
	\caption{2D XXZ model on $4\times 3$ square lattice with $J_x=J_y=1, J_z=0.8$. The spectrum decompositions of MPS ansatz with different bond dimensions are shown in the same figure. The training targets are ITES with $\beta=0.5$ (a) and $\beta=0.9$ (b). }
\label{fig:scatter}
\end{figure}

In addition, we also stack the spectral decomposition for a series of ansatzes in one figure to better illustrate the two-stage behavior for approximating ITES, see Fig. \ref{fig:scatter}.

\section{Other metrics for approximate imaginary-time evolved states}

\begin{figure}[t]\centering
	\includegraphics[width=0.6\textwidth]{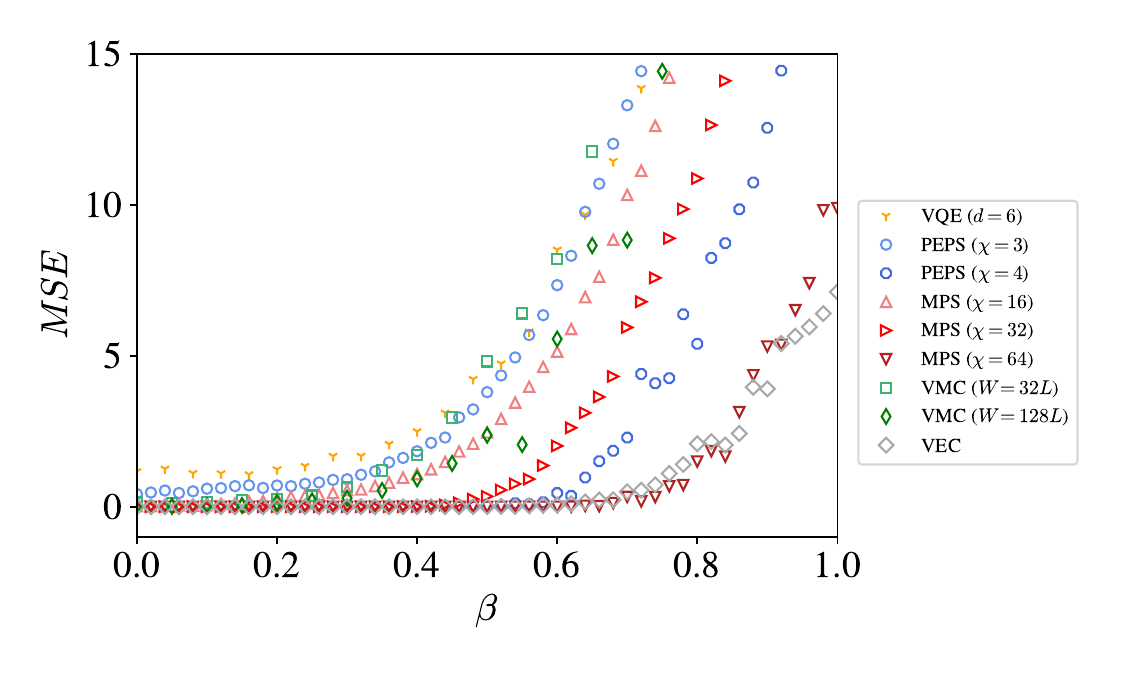}
	\caption{MSE for approximate ITES of different $\beta$s. The system is 2D XXZ model on $4\times 3$ square lattice with $J_x=J_y=1, J_z=0.8$. MSE begins to increase after $\beta^*$ for each ansatz.}
\label{fig:mse}
\end{figure}

In this section, we report some other metrics for approximate ITES as further evidence for the two-stage behavior. To diagnose the deviation from the exponential decay in the spectrum, we introduce mean squared error inspired by machine learning. Specifically, we construct MSE as $\frac{1}{2^L}\sum_{i=1}^{2^L}(y_i-y'_i)^2$, where the target sequence $y$ and approximate sequence $y'$ are spectral decomposition overlap $\log |c^0_i|^2$ for exact ITES and $\log |c_i|^2$ for approximate ITES, respectively. When MSE begins to deviate from 0 with increasing $\beta$, the training enters the low-temperature stage where $\tilde{\beta}<\beta$. The results of MSE are shown in Fig. \ref{fig:mse}.

\begin{figure}[t]\centering
	\includegraphics[width=0.6\textwidth]{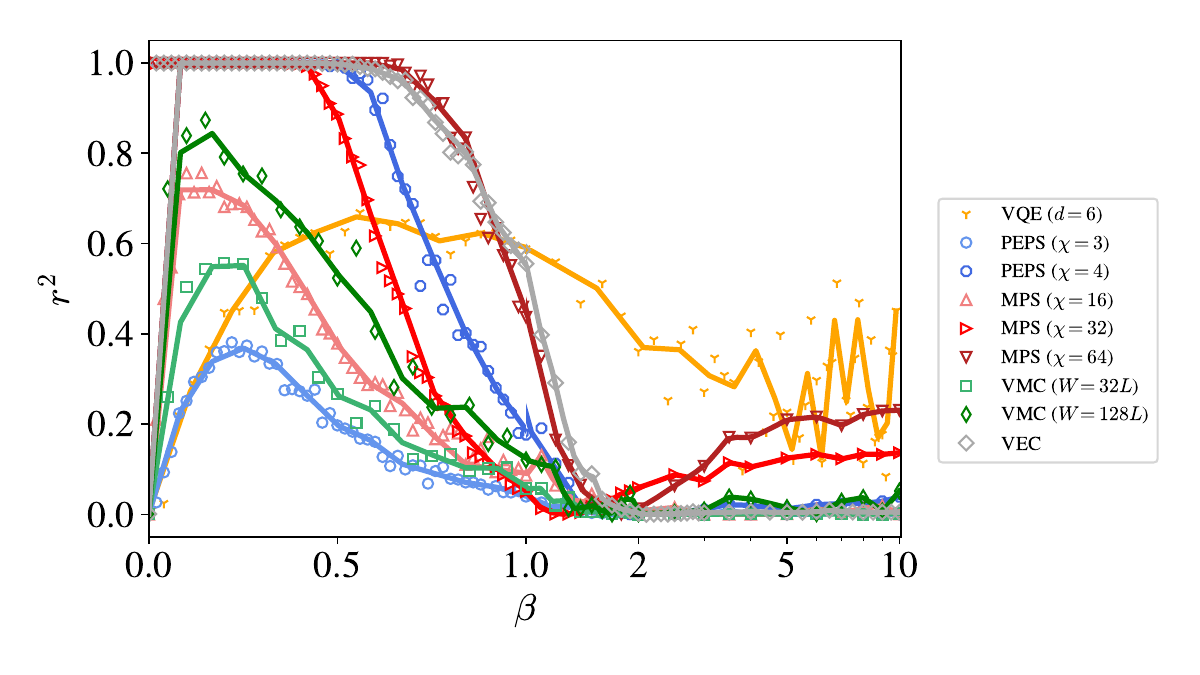}
	\caption{Squared Pearson coefficients for approximate ITES of different $\beta$s. The system is 2D XXZ model on $4\times 3$ square lattice with $J_x=J_y=1, J_z=0.8$. Pearson coefficient close to 1 indicates a good linear fitting.}
\label{fig:pearson}
\end{figure}

\begin{figure}[t]\centering
	\includegraphics[width=0.6\textwidth]{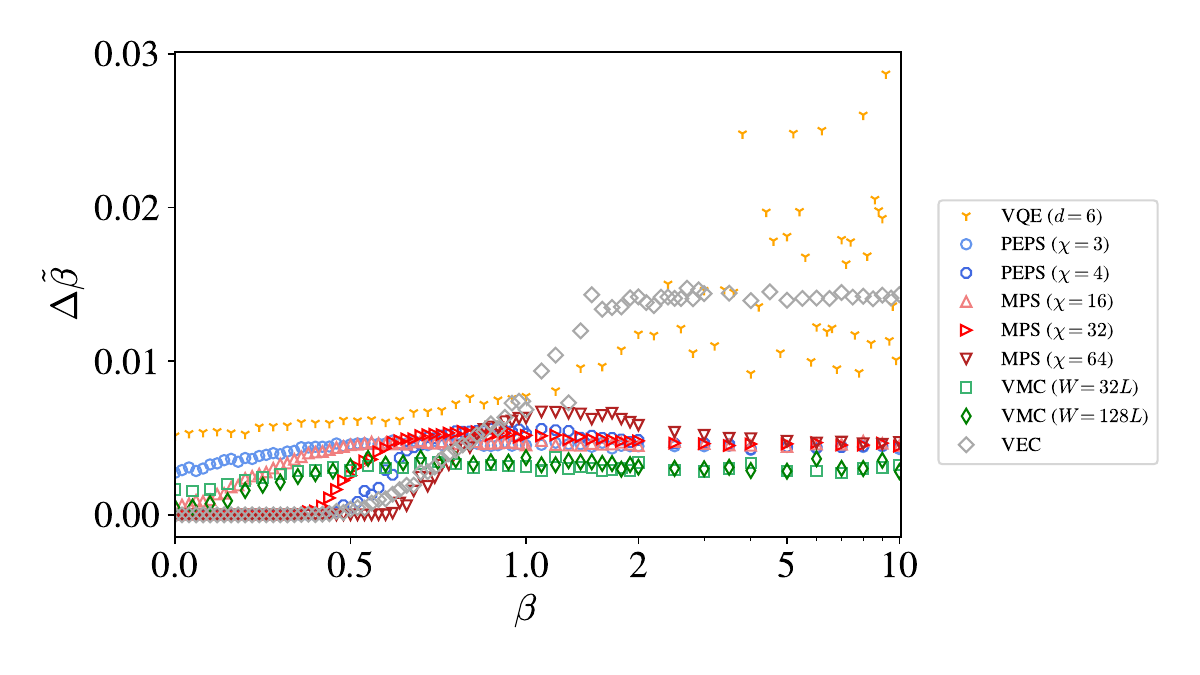}
	\caption{Uncertainty $\Delta \tilde{\beta}$ for approximate ITES of different $\beta$s. The system is 2D XXZ model on $4\times 3$ square lattice with $J_x=J_y=1, J_z=0.8$.}
\label{fig:deltabeta_ites}
\end{figure}

We also report the squared Pearson coefficients and the uncertainty $\Delta\tilde{\beta}$ in Fig. \ref{fig:pearson} and Fig. \ref{fig:deltabeta_ites}. Pearson coefficient close to 1 and uncertainty on $\tilde{\beta}$ close to 0 are both strong signals of good exponential fitting and the high-temperature stage. Notably, $\beta^*$ extracted from different metrics are consistent. It is an interesting future direction to explore whether universal critical behaviors can be identified around $\beta^*$ with these metrics.

\section{Expressiveness and trainability targeting imaginary-time evolved states}

\begin{figure}[t]\centering
	\includegraphics[width=0.48\textwidth]{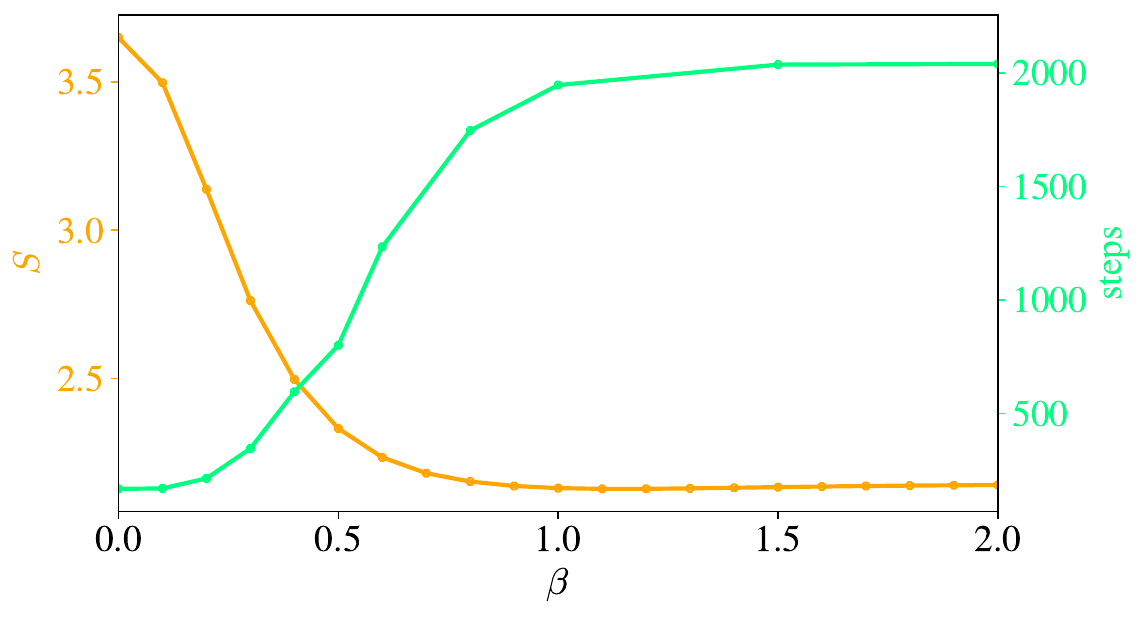}
	\caption{Demonstration on expressive capacity and optimization hardness for approximating ITES of different $\beta$. The MPS ansatz of bond dimension $\chi=32$ is employed. The system is 2D XXZ model on $4\times 3$ lattice with $J_x=J_y=1$, $J_z=0.8$. The orange line represents the entanglement entropy of the half system, related to the expressiveness required for accurate approximation. The green line represents the optimization steps required to reach a given fidelity accuracy. With lower temperature (higher $\beta$), the requirements on expressiveness are lower while the requirements on optimization are higher.}
\label{fig:tradeoff}
\end{figure}

In this section, we provide quantitative analysis for the change of required expressive capacity and optimization efforts with varying $\beta$ when approximating ITES. For the numerical simulation in this section, we use MPS ansatz with bond dimension $\chi=32$ to approximate ITES in 2D XXZ model on $4\times 3$ lattice with $J_x=J_y=1$, $J_z=0.8$. The optimization employs Adam optimizer with an exponential decay learning schedule (initial learning rate $10^{-3}$, and the learning rate halves every $1000$ step). We calculate the entanglement entropy for ITES of different $\beta$s, where the subsystem partition contains all sites on the left half. This quantity characterizes the expressive capacity required to accurately approximate the target states as higher entanglement requires larger bond dimensions to capture. We also conduct the optimization toward ITES and record the training steps required to reach a given infidelity threshold $10^{-7}$. Though the initialization and the optimizer are the same for different ITES, we find that the required number of training steps is much larger for low temperature (high $\beta$) targets. In other words, through the lens of ITES, we identify that the optimization becomes more difficult and slow for low temperature ITES near ground state manifold. The expressive capacity and optimization time required for approximating ITES of different temperatures are summarized in Fig. \ref{fig:tradeoff}. In this figure, we can see the evident change in expressiveness and optimization efforts required. 

Based on the above observations, we can explain the different trends of approximation accuracy with varying target $\beta$. For low-accuracy ansatzes, the bottleneck is at the expressive capacity to accurately capture the target state. Therefore, with larger $\beta$, the state is closer to the ground state with smaller entanglement entropy, resulting in looser expressiveness requirements and the performance of these low-accuracy ansatzes is improved. On the contrary, for high accuracy ansatzes, the ansatz is powerful enough to represent ITES of any temperature and the bottleneck is in optimization. As larger $\beta$ slows down the optimization, the fidelity obtained via a fixed number of training steps becomes worse. This understanding of the competition between expressiveness and optimization perfectly explains Fig. 4 (b) in the main text.

\begin{figure}[t]\centering
	\includegraphics[width=0.6\textwidth]{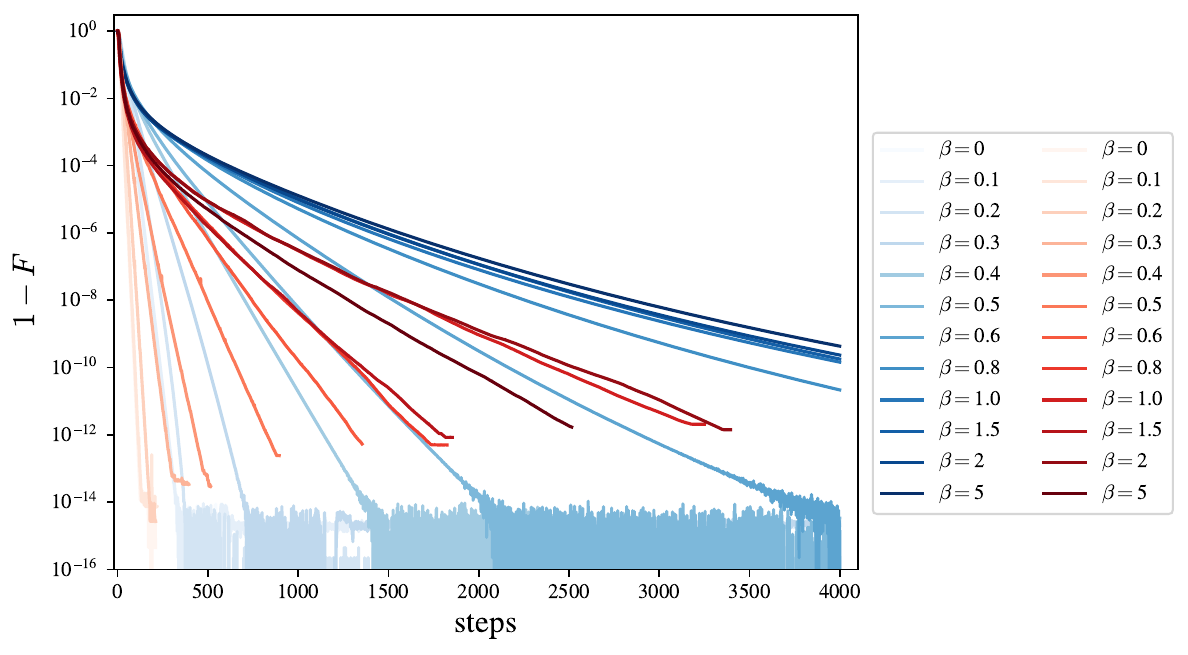}
	\caption{Fidelity during training for approximating ITES. The MPS ansatz of bond dimension $\chi=32$ is employed. The system is 2D XXZ model on $4\times 3$ lattice with $J_x=J_y=1$, $J_z=0.8$. The blue lines are training curves for Adam optimizer of identical hyperparameters while the red lines are training curves for L-BFGS optimizer of identical hyperparameters. L-BFGS greatly outperforms Adam optimizer, but the optimization hardness trend remains for low-temperature targets.}
\label{fig:bfgs}
\end{figure}

We also conduct the optimization using a more suitable optimizer L-BFGS (default hyperparameters provided by SciPy with function value and gradient tolerance $10^{-22}$ and max iteration $4000$), which greatly improves the fidelity for high-accuracy ansatzes in low-temperature regime (high $\beta$). This fact further supports the claim that the expressive capacity of these high-accuracy ansatzes is sufficient, the fidelity drop is mainly due to optimization hardness and insufficient training steps. The training dynamics for MPS ansatz ($\chi=32$) toward ITES with both Adam and L-BFGS optimizers are shown in Fig. \ref{fig:bfgs}.

\section{Hyperparameters for variational optimization}
For the optimization hyperparameters of each method, we use the following settings in Table. \ref{tab:hyper} unless explicitly specified.

\begin{table}[]
\caption{Hyperparameters for optimization with different ansatzes.}
\begin{tabular}{@{}cccc@{}}
\toprule
 ~Method~ & ~Optimizer~ & ~Initial learning rate~ & ~Learning schedule~  \\ \midrule
PEPS & Adam & $8\times 10^{-3}$ & constant \\
 MPS & Adam & $3\times 10^{-3}$ & exponential decay: halved each 1000 steps\\
 VMC & Adam & $1\times 10^{-3}$ & exponential decay (halved each 200 steps) at first 800 steps then constant\\ 
  VQE & Adam & $1\times 10^{-2}$ & exponential decay: halved each 2000 steps\\ 
 VEC & Adam & $2\times 10^{-3}$ & constant\\ 
 \bottomrule
\end{tabular}
\label{tab:hyper}
\end{table}

In terms of the number of training steps, we keep the optimization process until convergence when approximating ground states. When targeting ITES, the total training steps for each method are listed in Table \ref{tab:steps}.
\begin{table}[]
\caption{The number of training steps in ITES training.}
\begin{tabular}{@{}cc@{}}
\toprule
~Method~&~The number of training steps~\\ \midrule
PEPS &600 \\  
MPS & 400\\  
VMC &  3500\\ 
VQE & 2000\\  
VEC & 600\\ 
 \bottomrule
\end{tabular}
\label{tab:steps}
\end{table}

\end{document}